\documentclass[aps,prd,groupedaddress,preprint,eqsecnum,nofootinbib,showpacs]{revtex4}
\usepackage{graphicx,epsf,amssymb,amsbsy,amsfonts,amssymb,amsmath}
\usepackage[dvips]{color}

\newcommand{\beq}{\begin{equation}}
\newcommand{\eeq}{\end{equation}}
\newcommand{\be}{\begin{equation}}
\newcommand{\ee}{\end{equation}}
\newcommand{\beqa}{\begin{eqnarray}}
\newcommand{\eeqa}{\end{eqnarray}}
\newcommand{\beqar}{\begin{eqnarray*}}
\newcommand{\eeqar}{\end{eqnarray*}}
\newcommand{\bea}{\begin{eqnarray}}
\newcommand{\eea}{\end{eqnarray}}

\def\eq{\begin{equation}}
\def\eqae{\end{eqnarray}}
\def\eqa{\begin{eqnarray}}
\def\eqe{\end{equation}}
\def\be{\begin{equation}}
\def\ee{\end{equation}}
\def\bea{\begin{eqnarray}}
\def\eea{\end{eqnarray}}
\def\ba{\begin{array}}
\def\ea{\end{array}}
\def\bd{\begin{displaymath}}
\def\ed{\end{displaymath}}

\def\>{\rangle}
\def\<{\langle}
\def\Dsl{D \hskip-.6em \raise1pt\hbox{$ / $ } }
\def\to{\rightarrow}

\numberwithin{equation}{section}

\begin{document}

\setlength{\unitlength}{1mm}

\vspace{4mm}


\title{Amplitudes for Multiple M5 Branes}
\author{Bart{\l}omiej Czech${}^{a}$,
     Yu-tin Huang${}^{b}$, Moshe Rozali${}^{a}$ \\[10mm]
     {{${}^{a}${\it Department of Physics and Astronomy}\\
         {\it University of British Columbia}\\
         {\it 6224 Agricultural Road, Vancouver, BC V6T 1Z1,Canada}}\\[5mm]
         {${}^{b}${\it Department of Physics and Astronomy}\\
         {\it UCLA, Los Angeles, CA}\\
         {\it 90095-1547, USA}}\\[5mm]
$\null$\\
}
}

\vskip .3truecm

\begin{abstract}
We study $\mathcal{N}=(N,0)$ super-Poincar\'{e} invariant six-dimensional massless and five-dimensional massive on-shell amplitudes. We demonstrate that in six dimensions, all possible three-point amplitudes involving tensor multiplets are necessarily embedded in gravitational theories. For non-gravitational amplitudes we consider instead five-dimensional massive amplitudes with $\mathcal{N}=(2,0)$ supersymmetry, aiming at describing the world volume theory of multiple M5 branes compactified on $\mathcal{M}^{4,1}\times S^1$. We find non-gravitational amplitudes whose on-shell degrees of freedom are shown to match those of the massive particle states that arise from self-dual strings wrapping a circle. Along the way we find interesting hints of a fermionic symmetry in the $(2,0)$ theory, which accompanies the self-dual tensor gauge symmetry. We also discuss novel theories with $(3,0)$ and $(4,0)$ supersymmetry. 
\vskip .25truecm
\noindent 
\end{abstract}

\pacs{04.65.+e, 11.15.Bt, 11.30.Pb, 11.55.Bq \hspace{1cm}}

\maketitle

\tableofcontents

\section{Introduction and Discussion}

The existence of higher-dimensional field theories is one of the surprises discovered by indirect arguments using string dualities. Many such theories evade older arguments for the non-existence of fixed points above four dimensions by a combination of supersymmetry and strong coupling. Indeed, many of the theories which will be discussed here are not expected to have a tunable coupling constant and it is doubtful that they will have a Lagrangian description or whether such a Lagrangian could be a sensible starting point to approaching the quantum theory. This leaves an effective description of these higher dimensional theories an open challenge.

On the other hand, as long as the asymptotic states of the theory are well defined, one can always ask what is the S-matrix of the theory. Indeed, modern techniques of computing amplitudes, such as tree level recursion relations~\cite{Britto:2004ap,Britto:2005fq} and loop-level generalized unitarity methods~\cite{UnitarityMethod} do not require explicit knowledge of an underlying Lagrangian, or even its existence.\footnote{While in principle, one needs the underlying action to justify the validity of the recursion relations~\cite{ahk,cheung}, one can turn the argument around and ask under what conditions is the recursion valid and gain insight on the gauge algebra of the theory~\cite{Benincasa:2007xk}.} Rather, the amplitudes are constructed by requiring the correct factorization properties on poles and across branch cuts. Since the analytic structures of the S-matrix are well defined irrespective of whether the coupling is strong, one can always define the tree level on-shell amplitudes, i.e. the purely pole part of the S-matrix. Strong coupling is then simply the statement that those tree amplitudes cannot be considered as the leading approximation in a systematic perturbation expansion. The only ingredient needed for a construction of these tree level amplitudes via the tree level recursion relation is the three-point amplitude, which generically can be fixed by the symmetries of the theory and, in particular, the little group representation of the on-shell states.

While it has been shown that in the conformal phase the three- and four-point amplitude of the $\mathcal{N}=(2,0)$ self-dual tensor multiplet is trivial~\cite{huang}, recently it was shown in ref.~\cite{Samtleben:2011fj} that it is possible to write down an interacting action involving a self-dual tensor multiplet, albeit with reduced $\mathcal{N}=(1,0)$ supersymmetry.\footnote{A bosonic action involving self-dual tensors and vectors was given in~\cite{Chu:2011fd}. It can be shown that it is simply a special solution of the gauge system in~\cite{Samtleben:2011fj}. } While it is not clear whether these models can be successfully quantized, this construction hints at the possibility of gaining insight into chiral theories by putting them in the Coloumb phase, where the conformal invariance is broken and the fivebranes are separated.\footnote{More precisely, for the model proposed in~\cite{Samtleben:2011fj}, the interaction of the tensors is mediated by a vector whose quantization requires the theory to be on the broken phase since the usual kinematic term, $F^2$, appears with a scalar, $\phi F^2$.}

In this paper we would like to answer the following question: equipped with the on-shell asymptotic states, defined by the free theory, what is the maximum symmetry one can preserve to construct a non-trival amplitude. We will find that using unconstrainted on-shell variables, simple Lorentz invariance severly restricts the possible amplitudes that can be written down. Thus, we begin by relaxing both conformal and maximal supersymmetry constraints, and study the most general three-point amplitude with $\mathcal{N}=(N,0)$ super-Poincar\'{e} invariance involving self-dual tensor multiplets. This study is relevant to our original motivation, the (2,0) theories of multiple fivebranes, as well as to other theories with diverse matter content and at least (1,0) supersymmetry. 
  \subsection*{Three-Point Functions with (1,0) Supersymmetry}
As stated above, in this paper we sidestep the construction of a Lagrangian and its quantization and work directly with S-matrix elements. After reviewing the preliminaries in section II we devote section III to constructing all possible three-point amplitudes that involve $\mathcal{N}=(1,0)$ tensor, vector, and scalar multiplets. We find that super-Poincar\'{e} symmetry is a very strong constraint and one has only three possibilities: the scalar-scalar-vector, vector-vector-vector, and tensor-vector-vector multiplet interactions.  While the first two are simply the known super-Yang-Mills (sYM) three-point amplitude, the last one corresponds to the supersymmetrization of the following bosonic cubic term:
\eq
\epsilon^{\mu\nu\rho\sigma\tau\upsilon}B_{\mu\nu}F_{\rho\sigma}\bar{F}_{\tau\upsilon}\,
\label{result1}
\eqe
This term appears in the abelian tensor-vector interacting action proposed long ago by Siegel~\cite{Siegel:1983es}, as well as six-dimensional supergravity theories. This interaction also appeared in the action of~\cite{Samtleben:2011fj}, which predicted an additional vector-tensor-tensor amplitude, ruled out in our on-shell analysis. The resolution can be traced to the difficulty in quantizing, or equivalently defining asymptotic states, of the model. 

\subsection*{Three-Point Functions with Higher Spin and Higher Supersymmetry}
Extending the analysis for $\mathcal{N}=(1,0)$ to  $\mathcal{N}=(2,0)$, one can easily see that there are no three-self-dual tensor interactions with  $\mathcal{N}=(2,0)$ super-Poincar\'{e} invariance. The lack of $\mathcal{N}=(2,0)$ amplitudes for tensor multiplets alone is rather puzzling, as the only assumptions made were supersymmetry and the on-shell degrees of freedom. One way to relax the assumptions is to allow for the $\mathcal{N}=(2,0)$ tensor multiplet to interact with higher spin massless $\mathcal{N}=(2,0)$ multiplets. The introduction of higher spin states may be motivated as follows: from the M5 brane world-volume point of view, the M2 branes connecting separated M5 branes appear as self-dual strings. There is no controlled approximation by which one can deduce the properties of the theory of coincident fivebranes from these (so-called ``tensionless'') strings, but a natural guess is that on the Coulomb branch interactions of the tensor multiplets can be mediated by a tower of self-dual higher spin fields, presumably the excitations of the self-dual string.

We study this possibility in section IV. Allowing for higher spin (2,0) multiplets, one can write down a large class of $\mathcal{N}=(2,0)$ supersymmetric amplitudes involving various such multiplets. With the ansatz that the higher spin states of the new multiplets are self-dual, one obtains two $\mathcal{N}=(2,0)$ amplitudes involving the self-dual tensor multiplet. The first involves one tensor multiplet and two spin-3/2 fermion multiplets while the second involves two tensors and one graviton multiplet. The spin-3/2 multiplet contains a gravitino-like fermion that has the opposite chirality of the would-be gravitino for local $\mathcal{N}=(2,0)$ supersymmetry. The role of this mysterious spin-3/2 fermion is revealed once one realizes that the corresponding $\mathcal{N}=(2,0)$ amplitude is simply a truncation of the $\mathcal{N}=(2,2)$ supergravity amplitude. We thus see that all possible interactions involving self-dual tensor multiplets, either with reduced $\mathcal{N}=(1,0)$ supersymmetry or the full $\mathcal{N}=(2,0)$, can be embedded in supergravity amplitudes. 

While the $\mathcal{N}=(2,0)$ amplitudes we find are always a subset of $\mathcal{N}=(2,2)$ supergravity amplitudes, it is tempting to ask whether they define a consistent interacting theory by themselves. This question is motivated by the fact that $\mathcal{N}\leq2$ sYM amplitudes alone define a perfectly consistent interacting theory, despite being a supersymmetry truncation of $\mathcal{N}=4$ sYM amplitudes~\cite{Bern:2009xq,Lal:2009gn,HuangPeng}. One can answer this question by constructing the four-point amplitude via BCFW recursion and checking if it is consistent in all channels. A straightforward check reveals that it can only be made consistent if one allows gravitational couplings.

As a side result, the search for higher-spin interactions that involve only self-dual tensors lands us in an $\mathcal{N}=(3,0)$ super-Poincar\'{e} invariant three-point amplitude that describes the interaction of three self-dual rank 3 tensor multiplets, whose bosonic field content is 14 vectors and a self-dual 3rd-rank tensor. This multiplet can be viewed as a truncation of the $\mathcal{N}=(3,1)$ multiplet first proposed in refs.~\cite{Strathdee:1986jr,Townsend:1983xt}.

\subsection*{Three-Point Functions on $M^{{4,1}} \times S^{1}$}
The lack of non-gravitational amplitudes involving self-dual tensor multiplets in six dimensions implies that the asymptotic states we are working with are not the correct degrees of freedom. If we assume that the correct degrees of freedom are self-dual strings, then the previous analysis indicates that the approximation by higher-spin states is invalid. Another way of obtaining particle-like asymptotic states is to consider the theory on  $M^{{4,1}} \times S^{1}$, where the self-dual strings can wrap the circle and give a tower of massive KK modes in five dimensions~\cite{Lambert:2010iw}. Thus, one can instead study five-dimensional amplitudes with massive kinematics, which simply corresponds to breaking Lorentz covariance of six-dimensional massless kinematics down to the five-dimensional subgroup.

In section V, we consider massive five-dimensional amplitudes with $\mathcal{N}=(2,0)$ supersymmetry. Here, chirality is defined with respect to the massive little group, which is the same as the six-dimensional massless little group. Due to the breaking of manifest six-dimensional Lorentz symmetry, the constraint of super-Poincar\'{e} invariance admits new solutions. To constrain them, we require that upon reduction to four dimensions, the effect of the KK coupling must be such that the four-point one-loop amplitude of $\mathcal{N}=4$ sYM is modified simply by replacing the massless propagators of the box-integral to massive ones. This requirement is motivated by the study of Douglas~\cite{Douglas:2010iu}, which demonstrated that such interactions can give rise to an S-dualty invariant term in the effective action. In five dimensions, this constraint translates to the requirement that the one-loop four-point amplitude of the KK modes must be a box integral, i.e. there are no bubbles and  triangle integrals. This constrains the three-point amplitude to have mass dimensions $\leq1$. 

We present two solutions with $\mathcal{N}=(2,0)$ supersymmetry and mass dimension 1. From the little group structure of the amplitude one can deduce the participating multiplets which are shown to exactly match the expected KK spectrum coming from the wrapping of the self-dual string~\cite{Lambert:2010iw}. The presence of spin-3/2 particles in the higher spin multiplets requires local supersymmetry transformations which are particularly intriguing: they close to a gauge transformation of the self-dual tensor multiplet and do not relate the spin-3/2 fields to gravity. This suggests that the algebraic structure underlying the interacting theory on multiple coincident M5 branes has a novel fermionic symmetry, which is important for generalizing ordinary gauge invariance to non-Abelian tensor multiplets.

The purely self-dual tensor multiplet interaction is given by 
\eq
\mathcal{A}_3=\delta^5(P)\Delta(Q)\Delta(\hat{Q})\left( w_1^a\tilde{u}_{1a}\right),
\label{first}
\eqe
where the variables $u,w,\tilde{u}$ are defined for the special three-point kinematics and $(\Delta(Q), \Delta(\hat{Q}))$ are the chiral solutions to $\mathcal{N}=(2,0)$ supersymmetry constraints. Using six-dimensional BCFW recursion relations~\cite{cheung0,Dennen:2009vk}, we obtain a four-point amplitude for two massless and two KK tensor multiplets that has the correct factorization in both the $s$ and $t$ channels. It is given by
\eq
\mathcal{A}_4=\delta^5(P)\frac{\delta^4(Q)\delta^4(\hat{Q})}{s^{m}_{23}s^{0}_{14}}\,,
\eqe
where both fermionic delta functions are with respect to chiral supermomenta. Legs $2,3$ are the massless legs and $s^{m}_{23}=(p_2+p_3)^2+m^2,\;\;s^{0}_{14}=(p_1+p_4)^2$. Note that this is of the same form as the massive four-point amplitude in five-dimensional maximally supersymmetric YM, except that both supermomenta are chiral, compared to sYM where one is chiral and one anti-chiral.

The three-point amplitude in eq.~(\ref{first}) is only defined when one of the legs is massless. This is consistent with the recent proposal of effective bosonic action for multiple M5 branes in $D=5+1$~\cite{Ho:2011ni}, where the three-point interaction is mediated through massless modes.  The fact that the three-point amplitude can only be defined with a five-dimensional massless leg implies that such a particle description of the interactions cannot be lifted to six dimensions. Perhaps the six-dimensional interacting theory (on its Coulomb branch) must include string-like excitations as independent degrees of freedom.

In the massless limit, our amplitude reduces to the three-point amplitude of maximal sYM in five dimensions. Note that while the five-dimensional zero modes of the compactified six-dimensional self-dual tensor and vector are identical, their KK modes are distinct as they transform under different representations of the massive little group. Thus, our amplitude can be considered as the unique massive ``chiral" extension of the massless five-dimensional maximal sYM amplitude. Previous studies have indicated that the non-perturbative physics of the five-dimensional maximal sYM has information on the KK modes \cite{Rozali:1997cb} and may serve as the definition of the (2,0) theory (see~\cite{Lambert:2010iw, Douglas:2010iu} for a recent discussion). Our results indicate that the dynamics of the KK modes is simply a massive ``chiral" extension of the sYM amplitudes. 

An interesting application of the $\mathcal{N}=(2,0)$ self-dual tensor interaction is the construction of an interacting theory for the $\mathcal{N}=(4,0)$ multiplet. This multiplet was first proposed in~\cite{Hull:2000zn} and most recently studied in the context of unitary representations of the superconformal group~\cite{Chiodaroli:2011pp}. It was shown in~\cite{Hull:2000zn} that this theory dimensionally reduces to maximal supergravity in five dimensions, similar to the way the $\mathcal{N}=(2,0)$ theory reduces to maximal sYM. In other words, the $\mathcal{N}=(2,2)$ supergravity and the $\mathcal{N}=(4,0)$ theory correspond to two different ways of uplifting five-dimensional supergravity to six dimensions. While the interacting theory of the $\mathcal{N}=(4,0)$ multiplet is poorly understood and has no known M-theory embedding, one can construct a possible three-point amplitude. Following a similar analysis as for the self-dual tensor interaction, one immediately finds
\eq
\mathcal{A}_{3}^{N=(4,0)}=\delta^5(P)\left[\prod_{i=1}^4\Delta(Q^i)\right]\left( w_1^a\tilde{u}_{1a}\right)^2\,,
\eqe
where now one has a product of four different supermomentum delta functions. It is easy to see that this is simply a product of the $\mathcal{N}=(2,0)$ amplitude, and hence it reduces to maximal supergravity in the massless limit. Thus, in analogy, this massive amplitude corresponds to the unique chiral extension of the five-dimensional supergravity amplitude.

\subsection*{Conclusion}
In summary, we systematically study all possible three-point amplitudes involving self-dual tensor fields, subject to super-Poincar\'{e} invariance. Our analysis covers six-dimensional massless as well as five-dimensional massive kinematics, with the latter aiming at analyzing the interactions in the context of effective KK coupling. We find that six-dimensional three-point amplitudes involving self-dual tensors invariably can be embedded in a gravitational system. Non-gravitational couplings are kinematically allowed when we consider five-dimensional massive interactions, in which case we use constraints inspired by S-duality to fix the amplitude.  

We present two non-gravitational amplitudes in five dimensions. The on-shell degrees of freedom participating in the interaction can be directly matched to the BPS spectrum of a self-dual string wrapping a circle. One of the on-shell multiplets contains a spin-3/2 fermion, which would require a fermionic gauge symmetry. In the abelian context, we show that such a gauge transformation, which has the opposite chirality to a would-be gravitino, closes into the vector gauge symmetry of the self-dual tensor. The amplitude describing two-KK self-dual tensor coupling to a massless multiplet becomes maximal sYM in the massless limit. Thus, the interaction of the KK self-dual tensors is simply a massive chiral extension for the massless sYM amplitude, which we demonstrate at four-point as well. This result indicates that five-dimensional sYM captures part of the dynamics of the effective KK coupling. 

These results were derived by assuming the asymptotic states are the ones defined by the free theory. It is conceivable, for theories that do not have a tunable coupling, that these are the wrong degrees of freedom. It is interesting to explore interactions for a more general category of asymptotic states. One approach is to consider BPS states in six dimensions and construct massive amplitudes. These interactions should exist if we move to the Coulomb branch, where $N$ stacks of 5-branes are separated into an $M$- and $(M-N)$-stack. The analysis will be more involved as we do not have independent on-shell variables for massive six-dimensional kinematics.  

\subsection*{Organization}
In section II we give a brief review of the on-shell variables suitable for defining six-dimensional amplitudes, as well as the map between six-dimensional states and their four-dimensional massless descendants. In section III, after introducing variables for special three-point kinematics, we study supersymmetric amplitudes involving the self-dual tensor multiplet. In sections IV and V we generalize the search to include higher-spin interactions as well as higher supersymmetry. In section VI, we study massive amplitudes in five dimensions and find non-trivial three-point amplitudes, which describe the dynamics of the KK modes of the theory compactified on $\mathcal{M}^{4,1}\times S^1$.

\section{Preliminaries}
\subsection{On-shell $\mathcal{N}=(N,0)$ multiplets\label{multiplets}}
In this section we discuss general features for on-shell superfields for six-dimensional $\mathcal{N}=(N,0)$ supersymmetry. The on-shell degrees of freedom can be nicely packed into simple superfunctions by introducing Grassmann variables $\eta^{ma}$. Here the index $m$ runs from 1 to $N$ for an $\mathcal{N}=(N,0)$ multiplet and $a$ is an $SU(2)$ index, which can be identified as the chiral $SU(2)$ of the six-dimensional little group $SO(4)=SU(2)\times SU(2)$. These Grassmann variables can be identified as half of the fermionic components in the six-dimensional supertwistors, which we discuss in detail in the next subsection. The use of supertwistor~\cite{Ferber1977qx} variables to encode on-shell degrees of freedom for amplitudes is a straightforward generalization of the four-dimensional story~\cite{Nair:1988bq}. Note that the Grassmann variables do not carry R-symmetry indices as we have broken R-symmetry to maintain little goup covariance. 

For little-group covariant on-shell superspaces, self-CPT multiplets are given by scalar superfields. SUSY multiplets with non-maximal supersymmetry can then be obtained from the maximal ones by SUSY truncation, which entails integrating away or setting to zero some of the Grassmann variables in the on-shell superfield~\cite{Lal:2009gn,Bern:2009xq,HuangPeng}. We will first use the $\mathcal{N}=(2,0)$ and $(1,0)$ multiplets to illustrate these features.

The on-shell degrees of freedom of the $\mathcal{N}$=(2,0) tensor-multiplet can be packaged in a scalar superfield as follows~\cite{huang}:
\eqa
\nonumber\Phi(\eta^{a},\hat{\eta}^{a})^{\mathcal{N}=(2,0)}\!&=&\!\phi+\eta^{a}\chi_{a}+\hat{\eta}^{a}\chi'_{a}+\eta^{2}\phi'+\hat\eta^{2}\phi''+(\eta^{a}\hat\eta_{a})\phi'''+\eta_{(a}\hat\eta_{b)}B^{(ab)}\\
&+&\hat\eta^2\eta^{b}\bar\chi_{b}+\eta^{2}\hat{\eta}^{b}\bar\chi'_{b}+\eta^2\hat\eta^{2}\phi''''
\eqae
The component fields are the five scalars $(\phi,\phi',\phi'',\phi''',\phi'''')$, the three on-shell degrees of freedom of the self-dual tensor $B_{(ab)}$ transforming as a {\bf 3} under the $SU(2)$ little group, and the eight fermions.

There are three different on-shell multiplets for $\mathcal{N}$=(1,0) with spin$\leq$2: the scalar, tensor, and the vector multiplet. Since these are non-maximal SUSY, the multiplets are not self-CPT conjugate, which translates into the fact that the on-shell component fields are contained in two separate superfields instead of one as in the higher case. The scalar multiplet, which contains four scalars and four fermions, is given by: 
\eq
{\rm scalar}:\Phi(\eta^a)^{\mathcal{N}=(1,0)}=\phi+\eta^a\psi_a+(\eta^a\eta_a)\phi',\;\;\bar{\Phi}(\eta^a)^{\mathcal{N}=(1,0)}=\bar\phi'+\eta^a\bar\psi_a+(\eta^a\eta_a)\bar\phi\\
\label{scalar}
\eqe
The tensor multiplet contains one scalar, one self-dual tensor $B^{(ab)}$, and four fermions. It is represented by a pair of fermionic superfields transforming under the chiral $SU(2)$ little group:
\eq
{\rm tensor}:\Psi^b(\eta^a)=\chi^b+\eta^b\phi+\eta_aB^{(ab)}+(\eta^a\eta_a)\chi^{'b}
\label{tensor}
\eqe
The vector multiplet contains the four-component on-shell vector $G^a\,_{\dot a}$ and four fermions. It is represented by a pair of fermionic superfields transforming under the anti-chiral $SU(2)$ little group:
\eq
{\rm vector}:\Psi_{\dot a}(\eta^a)=\lambda_{\dot a}+\eta_aG^a\,_{\dot a}+(\eta^a\eta_a)\tilde\lambda_{\dot a}
\eqe
Note that the different multiplets are distinguished by their little group indices, which allows one to identify the multiplets involved in a given amplitude by inspecting the little group index structure. For example, a $6$-point amplitude carrying one chiral and one anti-chiral index involves one tensor, one vector, and $4$ scalar multiplets:  
\eq
(\mathcal{A}_6)_{a\dot{a}}\rightarrow \langle\Psi_a\Psi_{\dot{a}} \Phi\bar{\Phi}\Phi\bar{\Phi}\rangle
\eqe

From the above, one can easily see that the $\mathcal{N}=(1,0)$ tensor multiplet can be obtained by integrating away one of the $\hat{\eta}$'s from the $\mathcal{N}=(2,0)$ on-shell superfield, and setting the remaining $\hat{\eta}$'s to zero:
\eq
\Psi^b=\left.\left[ \int d\hat{\eta}_{b}\,\Phi(\eta,\hat{\eta})^{\mathcal{N}=(2,0)}\right]\right|_{\hat{\eta}=0}
\eqe
The $\mathcal{N}=(1,0)$ scalar multiplets, on the other hand, can be obtained by integrating away the two $\hat{\eta}$'s and setting all $\hat{\eta}$'s to zero:
\eq
\Phi^{\mathcal{N}=(1,0)}=\Phi^{\mathcal{N}=(2,0)}|_{\hat{\eta}=0},\quad\quad\bar{\Phi}^{\mathcal{N}=(1,0)}=\int d^2\hat{\eta}\, \Phi^{\mathcal{N}=(2,0)}
\eqe
The $\mathcal{N}=(1,0)$ vector multiplet can similarly be obtained from the maximal $\mathcal{N}=(1,1)$ vector multiplet~\cite{Dennen:2009vk,HuangPeng}. This embedding of lower supersymmetric multiplets is a very useful procedure for obtaining lower supersymmetric amplitudes from higher ones~\cite{Bern:2009xq,HuangPeng}.

The above discussion can be easily generalized to $\mathcal{N}=(N,0)$ with $N\geq3$. These are multiplets that include higher spins. As examples, we consider the $\mathcal{N}=(4,0)$ and $\mathcal{N}=(3,1)$ multiplet, whose interaction will be discussed in this paper. The  $\mathcal{N}=(4,0)$ multiplet was first discussed in detail in~\cite{Hull:2000zn} as a candidate theory for conformal supergravity in six dimensions and also more recently in the context of unitary representations of the superconformal group in~\cite{Chiodaroli:2011pp}. The $\mathcal{N}=(3,1)$ multiplet was first suggested in ref.~\cite{Strathdee:1986jr,Townsend:1983xt}. Both the $\mathcal{N}=(3,1)$ and the $\mathcal{N}=(4,0)$ multiplet, while having different field content in six dimensions due to chirality, reduce to maximal supergravity in five dimensions~\cite{Hull:2000zn}. 

The bosonic field content of the maximal $\mathcal{N}=(3,1)$ multiplet contains 28 scalars $\phi$, 14 vectors $A^{\mu}$, 12 self-dual two forms $B^{\mu\nu}$ and 1 self-dual rank 3 tensor $D^{\mu\nu\rho}$. While the self-dual two-form transforms as a ({\bf 3},{\bf 1}) under the little group $SU(2)\times SU(2)$, the self-dual rank 3 tensor transforms as a ({\bf 4},{\bf 2}) and can be denoted as $D^{(abc),\dot{a}}$. The symmetry of the rank 3 tensor is~\cite{Hull:2000zn}
\eq
D_{\mu\nu\rho}=-D_{\nu\mu\rho},\;\;D_{[\mu\nu\rho]}=0\,,
\eqe
with the abelian gauge transformation given by  
\eq
\delta D_{\mu\nu\rho}=\partial_{[\mu}\Lambda_{\nu]\rho}-\partial_{[\mu}\Lambda_{\nu\rho]}\,.
\eqe
The field strength is defined with double derivatives and is given by
\eq
I_{\mu\nu\rho\sigma\tau}=\partial_{[\mu}D_{\nu\rho][\sigma,\tau]}
\eqe
and the self-duality condition is defined as 
\eq
I_{\mu\nu\rho\sigma\tau}=\frac{1}{3!}\epsilon_{\mu\nu\rho\omega\upsilon\chi}I^{\omega\upsilon\chi}\,_{\sigma\tau}
\eqe
The on-shell states can be neatly packed into a Lorentz scalar superfield $\Phi(\eta^m_a,\tilde{\eta}_{\dot{a}})$, where $m,n,p\in\{1,2,3\}$, with the 128 bosonic states appearing with even degrees of $\eta,\tilde{\eta}$, as:   
\begin{center}
    \begin{tabular}{ | p{0.3cm}  | p{5.0cm} | p{3.5cm} | p{3.4cm} | l |}
    \hline
    $d$ & $\phi$ & $A^{a\dot{a}}$ & $B^{(ab)}$& $D^{(abc),\dot{a}}$ \\ \hline
    2&$(\eta^m)^{2}\,[3]$,\,$\eta^{mb}\eta^{n}_{\phantom{n}b}\,[3]$,\,$\tilde{\eta}^2\,[1]$ & $\eta_a^{m}\tilde\eta_{\dot{a}}\,[3]$ & $\eta_{(a}^{m}\eta_{b)}^{n}\,[3]$ & [0] \\ \hline
   4& $(\eta^{m})^2(\eta^n)^{2}\,[3]$,\,$(\eta^{m})^2\tilde\eta^{2}\,[3]$,

$\eta^{mb}\eta^{n}_{\phantom{n}b}\tilde\eta^{2}\,[3]$,\,\,$\eta^{mb}\eta^{n}_{\phantom{n}b}(\eta^{p})^2\,[3]$ & $\eta^{mb}\eta^{n}_{\phantom{n}b}\eta_{\phantom{m}a}^{m}\tilde\eta_{\dot{a}}\,[2]$,\,\,\,

$(\eta^{p})^2\eta_{\phantom{m}a}^{m}\tilde\eta_{\dot{a}}\,[6]$ 
& $\tilde{\eta}^2\eta_{\phantom{m}(a}^{m}\eta_{\phantom{n}b)}^{n}\,[3]$,
   
   \,$(\eta^{p})^2\eta_{\phantom{m}(a}^{m}\eta_{\phantom{n}b)}^{n}\,[3]$ 
& $\eta^{m}_{\phantom{m}(a}\eta_{\phantom{n}b}^{n}\eta_{\phantom{p}c)}^{p}\tilde{\eta}^{\dot{a}}\,[1]$ \\\hline
    6& $(\eta^{m})^2(\eta^{n})^2(\eta^{p})^2\,[1]$,
     \,$(\eta^{m})^2(\eta^{n})^2\tilde\eta^{2}\,[3]$,\,

    $\eta^{mb}\eta^{n}_{\phantom{n}b}(\eta^{p})^2\tilde\eta^{2}\,[3]$ 
& $(\eta^{n})^2(\eta^{p})^2\eta_{\phantom{m}a}^{m}\tilde\eta_{\dot{a}}\,[3]$ 
& $(\eta^{p})^2\tilde{\eta}^2\eta_{\phantom{m}(a}^{m}\eta_{\phantom{n}b)}^{n}\,[3]$ 
&  [0] \\
        \hline
    \end{tabular}
\end{center}
Here $(\eta^{m})^2\equiv \eta^{ma}\eta^m_{\phantom{m}a}$, the degree $d$ labels the total power of $\eta,\tilde{\eta}$'s and the number of fields is denoted in the square brackets. Note that the scalars also include those appearing in the superfield expansion at degree 0 and 8, which are not listed above for brevity. A similar discussion can be applied to the $\mathcal{N}=(4,0)$ multiplet. From the component expansion of the scalar superfield  $\Phi(\eta^m_{\phantom{m}a})$, where now $m=1,\cdot\cdot,4$, one sees that the bosonic field content contains 42 scalars, 27 self-dual two-forms and one rank 4 self-dual tensor that transforms as a ({\bf 5},{\bf 1}) under $SU(2)\times SU(2)$.

One can obtain $\mathcal{N}=(3,0)$ multiplets by SUSY truncation of the $\mathcal{N}=(3,1)$ multiplet as discussed previously. One integrates away one $\tilde{\eta}$ and sets the remaining $\tilde\eta$ to zero, which gives two fermionic superfields that form a doublet under the anti-chiral $SU(2)$ little group, $\Psi^{\dot{a}}(\eta^m_{\phantom{m}a})$ with $m=1,2,3$. The bosonic components are given as:
 \begin{center}
    \begin{tabular}{ | l  | l | p{4.8cm} | l | p{2.5cm} |}
    \hline
    $d$ & $\phi$ & $A^{\mu}$ & $B^{\mu\nu}$& $D^{\mu\nu\rho}$ \\ \hline
    1&$[0]$ & $\eta_{\phantom{m}a}^{m}\,[3]$ & [0] & [0] \\ \hline
   3& [0] & $\eta^{mb}\eta^{n}_{\phantom{n}b}\eta_{\phantom{m}a}^{m}\,[2]$,\,$(\eta^{p})^2\eta_{\phantom{m}a}^{m}\,[6]$ & [0] & $\eta^{m}_{\phantom{m}(a}\eta_{\phantom{n}b}^{n}\eta_{\phantom{p}c}^{p}\,[1]$ \\\hline
    5& [0] & $(\eta^{n})^2(\eta^{p})^2\eta_{\phantom{m}a}^{m}\,[3]$ & [0] & 
   [0] \\
        \hline
    \end{tabular}
\end{center}
The bosonic fields of this $\mathcal{N}=(3,0)$ multiplet consist of 14 vectors and 1 self-dual rank 3 tensor, a total of 64 states. The on-shell states of the $\mathcal{N}=(3,0)$ multiplet can be obtained as a tensor product of the $\mathcal{N}=(2,0)$ and the $\mathcal{N}=(1,0)$ vector multiplet. Note that the fermionic states of the above $\mathcal{N}=(3,0)$ and $\mathcal{N}=(3,1)$ multiplets contain gravitino-like spinors $\psi_{(ab)}\,^{\dot{a}}$, which are spin-$3/2$ particles that transform as a $(\mathbf{3},\mathbf{2})$ under the little group and have opposite chirality to that of a gravitino. To see this, observe that in a chiral multiplet the gravitino would have the opposite chirality to the supercharge since 
\eq
[\epsilon_AQ^A,\; (e_\mu)_{BC}]=\epsilon_{[B}(\psi_\mu)_{C]},
\eqe 
where we have written the local Lorentz indices in the $SU^*(4)$ spinor representation. In terms of little group representation, the gravitino is written as $\psi_{(\dot{a}\dot{b})}\,^a$ and transforms as a $(\mathbf{2},\mathbf{3})$. Thus, we see that our gravitino-like spinor indeed has the opposite chirality to the gravitino. The opposite chirality results in the fact that although the free theory also has local SUSY, the anti-commutation of the local SUSY produces the vector gauge symmetry of self-dual tensors instead of gravitons.

One can obtain other $\mathcal{N}=(3,0)$ multiplets by performing similar SUSY truncations on the $\mathcal{N}=(4,0)$ mutiplet instead. This will give either a chiral fermionic or a pair of scalar superfields, depending on whether one integrates one or two powers of $\eta^m_{\phantom{m}a}$ with $m=4$. In fact, this simply amounts to choosing a different Clifford vacuum for the construction of on-shell SUSY representations. One can easily obtain higher spin multiplets by choosing the vacuum to be higher rank in little group indices or, equivalently, allowing the on-shell superfields to transform as higher-rank tensors.

Finally, the Grassmann degree of the superamplitude is fixed by the R-symmetry constraint. For a generic $\mathcal{N}=(N,0)$ multiplet the R-symmetry is $Sp(2N)$, which contains as a subgroup $Sp(2)^N$. To be invariant under each $Sp(2)$ one requires the $n$-point amplitude to vanish under the following $U(1)$ generator: 
\eq
\mathcal{G}^m_{U(1)}\equiv \sum_i^n\left(\eta_i^{ma}\frac{\partial}{\partial \eta_i^{ma}}-1\right)
\eqe 
Here $i$ labels the external legs and there is one $\mathcal{G}_{U(1)}$ for each copy of $Sp(2)$. The explicit form of $\mathcal{G}^m_{U(1)}$ can be derived from the superconformal algebra discussed in Sec.~\ref{SC}. Thus, the $Sp(2)^N$ R-symmetry subgroup requires the $n$-point amplitude of interacting $\mathcal{N}=(N,0)$ multiplets to be of Grassman degree $n\cdot N$.

\subsection{Superconformal generators and on-shell variables\label{SC}}
In this subsection, we give a brief review of six-dimensional supertwistors. Supertwistors are a convenient starting point for developing the spinor helicity formalism since the variables can be straightforwardly used as the on-shell variables for the amplitude and form a representation of the superconformal algebra. This allows one to construct amplitudes out of building blocks invariant with respect to the symmetry generators, which the amplitude is expected to respect. 

For the six-dimensional superconformal group, supertwistors are in the fundamental representation of the supergroup $OSp^*(8|2N)$, where the $^*$ stands for psuedoreal. The supertwistors are given by
$$\mathcal{Z}^{\mathcal{M}a}=\left(\begin{array}{c}Z^{\mu a}\\ \eta^{Ia}\end{array}\right),$$
where $\mu=1,\cdot\cdot,8$ are $SO(2,6)=SO^*(8)$ indices, $I=1,\cdot\cdot,2N$ are $Sp(2N)$ R-indices and finally $a=1,2$ are the $SU(2)$ little group indices. 

One can construct the superconformal generators using the above supertwistors by identifying the superconformal generators as 
$$G^{\mathcal{M}}_{\phantom{\mathcal{M}}\mathcal{N}}
=\mathcal{Z}^{\mathcal{M}a}\mathcal{Z}_{\mathcal{N}a}.$$
To generate the group algebra, one only has to take into account that the supertwistors are self-canonical conjugate, which means that only half the components of a supertwistor are independent:
\be
[\mathcal{Z}^{\mathcal{M}a},\mathcal{Z}^b_{\mathcal{N}}\}=\delta^{\mathcal{M}}_{\phantom{\mathcal{M}}\mathcal{N}}\epsilon^{ab}
\quad\Rightarrow\quad 
[Z^{\mu a},Z^{\nu b}]=\eta^{\mu\nu}\epsilon^{ab} \quad\quad
\{\eta^{Ia}, \eta^{Jb}\}=\epsilon^{ab}\Omega^{IJ}
\ee
Here $\eta^{\mu\nu},\;\epsilon^{ab},\;\Omega^{IJ}$ are the $SO^*(8)$, $SU(2)$, and $USp(2N)$ metrics, respectively.

To separate the independent pieces, one recombines the 8 $SO^*(8)$ bosonic twistors into the {\bf 4} and ${\bf \bar{4}}$ of $SU^*(4)$, which is the covering group of the six-dimensional Lorentz group $SO(5,1)$:
\eq
Z^{\mu a}=\left(\begin{array}{c}\lambda^{Aa} \\ \mu^a_{A}\end{array}\right)
\eqe 
The self-canonical conjugate relation becomes $[\lambda^{Aa},\mu^b_{B}]=\delta^{A}_B\epsilon^{ab}$. Thus, one can set 
\eq
\mu^a_{A}=\frac{\partial}{\partial \lambda^{Aa}}\,.
\eqe
For the fermionic twistors we note that $Sp(2N)$, the R-symmetry group for the $(N,0)$ theory, can be broken down to $Sp(2)^{N}=SU(2)^{N}$. Thus, the $2N$ components $\eta^{Ia}$ can be regrouped as $N$ $Sp(2)$ spinors 
\eq
\eta^{I a}=\left(\begin{array}{c}\eta^{ma+} \\ \eta^{ma-}\end{array}\right),
\eqe 
where $m=1,\cdot\cdot,N$ labels copies of $Sp(2)$ and the $\pm$ indicates its charge under the $U(1)$ subgroup of each $Sp(2)=SU(2)$, i.e. the $J_z$ component in each $SU(2)$. The self-conjugate relation $\{\eta^{Ia},\eta^{Jb}\}=\epsilon^{ab}\,\Omega^{IJ}$ now becomes:
\eq
\{\eta^{ma-},\eta^{nb+}\}=\epsilon^{ab}\epsilon^{-+}\eta^{mn},\;\;\{\eta^{ma-},\eta^{nb-}\}=\{\eta^{ma+},\eta^{nb+}\}=0
\eqe
Thus, one can identify:
\eq
\eta^{m-}=\frac{\partial}{\partial \eta^{m+}}
\eqe

In the end, the independent variables are ($\lambda^{Aa},\eta^{ma}$); for simplicity from now on we drop the superscript $+$ on the $\eta$'s. The superconformal generators can be straightforwardly written as:
\eq
\begin{array}{rclp{1cm}l}
P^{AB}&=& \lambda^{Aa}\lambda^B_{a} & & [6]\\
Q^{mA-}&=&\lambda^{Aa}\frac{\partial}{\partial \eta^{ma}},\quad Q^{mA+}=\lambda^{Aa}\eta^{m}_a & & [8\mathcal{N}]\\
M^A_{\phantom{A}B}&=&\lambda^{Aa}\frac{\partial}{\partial\lambda^{Ba}}-\frac{\delta^A_{\phantom{A}B}}{4}\lambda^{Ca}\frac{\partial}{\partial\lambda^{Ca}}& & [15=16-1]\\
D&=&\frac{1}{2}\lambda^{Aa}\frac{\partial}{\partial \lambda^{Aa}}+2 & & [1]\\
S_{A}^{m+}&=&\eta^{ma}\frac{\partial}{\partial \lambda^{Aa}},\quad S_{mA}^{-}=\frac{\partial}{\partial \lambda^{Aa}}\frac{\partial}{\partial \eta^{m}_a}& & [8\mathcal{N}]\\
K_{AB}&=&\frac{\partial}{\partial \lambda^{Aa}}\frac{\partial}{\partial \lambda^{B}_a}& & [6]
\end{array}
\label{theman}
\eqe
Defining the amplitude as a function of the variables ($\lambda^{Aa},\eta^{ma}$), we can impose whichever symmetry is desired by requiring that the corresponding generators in eq.~(\ref{theman}) vanish on the amplitude.

From eq.~(\ref{theman}) we see that the six-dimensional massless momenta are represented as~\cite{cheung0,Boels:2009bv,Dennen:2009vk,Bern:2010qa}
\eq
p_i^{AB}=\lambda_i^{Aa}\lambda^{B}_{ia}.
\eqe 
It is anti-symmetric in the two $SU^*(4)$ indices due to the contraction of the $SU(2)$ index. We can define the anti-chiral spinors as:
\eq
\tilde{p}_{iAB}=\tilde\lambda_{iA\dot{a}}\tilde{\lambda}_{iB}^{\dot{a}}=\frac{1}{2}\epsilon_{ABCD}p_i^{CD}
\eqe
The six-dimensional Lorentz invariant spinor inner products are
\eq
\langle i_a|j_{\dot{b}}]=\lambda_i^{Aa}\tilde\lambda_{jA\dot{b}}.
\eqe 
The vector inner product can be expressed in terms of the spinor inner products as
\eq
p_i^\mu p_{j\mu}=\frac{1}{4}p_i^{BA}p_{jAB}=\frac{1}{2}\det \langle i_a|j_{\dot{a}}].
\label{det}
\eqe
\subsection{Three-point kinematics}
At three points we have $p_i\cdot p_j=0$, so from eq.~(\ref{det}) the Lorentz invariant spinor product $\langle i_a|j_{\dot{a}}]$ is rank one. It is then more suitable to follow \cite{cheung0} and introduce $SU(2)$ spinors $u^a,\tilde{u}^{\dot a}$
\eqa
\nonumber\langle1_a|2_{\dot{b}}]=u_{1a}\tilde{u}_{2\dot{b}}\quad&\quad\langle2_a|1_{\dot{b}}]=-u_{2a}\tilde{u}_{1\dot{b}}\\
\langle2_a|3_{\dot{b}}]=u_{2a}\tilde{u}_{3\dot{b}}\quad&\quad\langle1_a|3_{\dot{b}}]=-u_{1a}\tilde{u}_{3\dot{b}} \label{defu}\\
\langle3_a|1_{\dot{b}}]=u_{3a}\tilde{u}_{1\dot{b}}\quad&\quad\langle3_a|2_{\dot{b}}]=-u_{3a}\tilde{u}_{2\dot{b}} \nonumber
\label{not}
\eqae
along with their pseudo-inverses $w_{ia}$ (with similarly definitions for $\tilde{w}_{i\dot{a}}$):
\eq
u_{ia}w_{ib}-u_{ib}w_{ia}=\epsilon_{ab}\quad\Rightarrow\quad u_i^a w_{ia}=1\,.
\label{Invert}
\eqe
Note that the above definitions are invariant under the shift:
\eq
 w_{ia} \to w_{ia} + b_i u_{ia}\,
\label{bShift}
\eqe
This $b$-shift ambiguity implies that amplitudes must be invariant under eq.~(\ref{bShift}).

These variables satisfy identities that follow from momentum conservation:
\begin{align}
\nonumber\tilde{\lambda}_{1B\dot{b}}(p_1+p_2+p_3)^{AB}=\tilde{u}_{1\dot{b}}(u_{3a}\lambda_3^{Aa} -u_{2a}\lambda_2^{Aa})=0 \quad \Rightarrow \quad 
& u_{3a}\lambda_3^{Aa}=u_{2a}\lambda_2^{Aa}=u_{1a}\lambda_1^{Aa} \\
\Rightarrow \quad & \tilde u_{3}^{\dot a}\tilde\lambda_{3A\dot{a}}=\tilde u_{1}^{\dot a}\tilde\lambda_{1A\dot{a}}=\tilde u_{2}^{\dot a}\tilde\lambda_{2A\dot{a}}
\label{Ueq}
\end{align}
Momentum conservation combined with eq.~(\ref{Invert}) also implies: 
\begin{equation}
\sum_{i=1}^3\lambda_i^{Aa}\epsilon_{ab}\lambda_i^{Bb}=\sum_{i=1}^3\lambda_i^{Aa}(u_{ia}w_{ib}-u_{ib}w_{ia})\lambda_i^{Bb}=u_{1a} \lambda_1^{a[A}
\sum_{i=1}^3\lambda_i^{B]b}w_{ib}=0 \quad\Rightarrow\quad\sum_{i=1}^3\lambda_i^{Bb}w_{ib}=0 \label{Weq}
\end{equation}
The second equality follows from eq.~(\ref{Ueq}). Note that the final constraint reduces the $b$-shift freedom to $w_i\rightarrow w_i + b_iu_i$ with $\sum_i^3b_i=0$. Here and in the following the $SU(2)$ indices are raised and lowered with $\epsilon_{ab}$ and $\epsilon_{\dot{a}\dot{b}}$; our conventions are given in Appendix~\ref{conventions}.

At three points, with full six-dimensional Lorentz invariance, one can always choose the three vectors to lie in a four-dimensional subspace and obtain massless four-dimensional kinematics. In this particular frame, the six-dimensional spinors can be written as~\cite{cheung0,Bern:2010qa}
\eq
\lambda^{A}_a=\left(\begin{array}{cc}0 & \lambda_\alpha^{(4)} \\ \tilde\lambda^{(4)\dot\alpha} & 0\end{array}\right),\quad\quad \tilde{\lambda}_{A\dot{a}}=\left(\begin{array}{cc}0 & \lambda^{(4)\alpha} \\ -\tilde\lambda^{(4)}_{\dot\alpha} & 0\end{array}\right),
\label{spinor}
\eqe
where we have used the superscript $(4)$ to denote four-dimensional variables. The spinor products are then given by\footnote{Here the conventions are $\langle ij\rangle\equiv\lambda_i^\alpha\lambda_{j\alpha}$ and $[ ij]\equiv\tilde\lambda_{i\dot{\alpha}}\tilde\lambda_{j}^{\dot{\alpha}}$.} 
\eq
\langle i_a|j_{\dot{a}}]=\left(\begin{array}{cc}[ij] & 0 \\0 & -\langle ij\rangle\end{array}\right).
\eqe
For three-point kinematics, we have the extra condition that $[ij]\langle ij\rangle=0$. With complex kinematics, one can choose either $\langle ij\rangle=0$ or $[ij]=0$. For $\langle ij\rangle=0$, one can write explicit solutions for $u_i, \tilde{u}_i$ using four-dimensional spinor variables:
\begin{align}
u_{1a} = ( \, 0, \quad \langle 2 3 \rangle^{-1} \, ) \quad & \quad \tilde{u}_{1\dot{b}} = (\, 0, \quad -\langle 1 2 \rangle \langle 3 1 \rangle \, ) \nonumber \\
u_{2a} = ( \, 0, \quad \langle 3 1 \rangle^{-1} \, ) \quad & \quad \tilde{u}_{2\dot{b}} = (\, 0, \quad -\langle 1 2 \rangle \langle 2 3 \rangle \, ) \label{expl6} \\
u_{3a} = ( \, 0, \quad \langle 1 2 \rangle^{-1} \, ) \quad & \quad \tilde{u}_{3\dot{b}} = (\, 0, \quad -\langle 2 3 \rangle \langle 3 1 \rangle \, ) \nonumber
\end{align}
The pseudoinverses $w$ and $\tilde{w}$ take the form:
\begin{align}
w_{1a} = ( \,  -\langle 2 3 \rangle,\quad 0 \, )  \quad & \quad 
\tilde{w}_{1\dot{b}} = (\, \langle 1 2 \rangle^{-1} \langle 3 1 \rangle^{-1},\quad 0 \, ) 
\nonumber \\
w_{2a} = ( \,  -\langle 3 1 \rangle,\quad 0 \, ) \quad & \quad
\tilde{w}_{2\dot{b}} = (\, \langle 1 2 \rangle^{-1} \langle 2 3 \rangle^{-1},\quad 0 \, ) 
\label{explw} \\
w_{3a} = ( \,  -\langle 1 2 \rangle,\quad 0 \, ) \quad & \quad
\tilde{w}_{3\dot{b}} = (\,\langle 2 3 \rangle^{-1} \langle 3 1 \rangle^{-1},\quad 0  \, )
\nonumber
\end{align}
These solutions are defined only up to the rescaling:
\eq
u\rightarrow \alpha u,\;\;\tilde{u}\rightarrow \alpha^{-1} \tilde{u}, \;\; w\rightarrow \alpha^{-1} w,\;\;\tilde{w}\rightarrow \alpha \tilde{w}
\label{alpha}
\eqe 
This scaling ambiguity is inherent in the definition (\ref{defu}). Thus, requiring that the three-point superamplitude be invariant under this rescaling is necessary for six-dimensional Lorentz invariance. The solutions for $\langle ij\rangle=0$ are given in Appendix~\ref{SquareSol}.

\subsection{Dimensional reduction to $D=4$\label{Map}}
Once the three-point amplitudes are found using on-shell variables, one can reconstruct the corresponding interaction terms by studying components and symmetries of the amplitude. The results can be verified by using the explicit form of polarization vectors or tensors written in terms of on-shell variables.

An easier way is to dimensionally reduce the proposed interaction to four dimensions and compare the result with the dimensional reduction of the three-point amplitude. For the amplitude, this is done by choosing the four-dimensional subspace to include that spanned by the three momenta. This will be referred to as the massless reduction. For this purpose it will be useful to map the massless states in six dimensions to four-dimensional ones. 

We first project the six-dimensional on-shell states to four-dimensional massless ones. For consistency, the dimensional reduction of the $\eta^a$ follows from the reduction of the bosonic spinors given in eq.~(\ref{spinor}):
\eq
6D: \eta_a\rightarrow 4D:\left(\begin{array}{c} \zeta^1 \\ \bar{\zeta}_2\end{array}\right)
\eqe
Here we have used $\zeta$'s to denote the four-dimensional Grassmann variables. 
The indices $1,2$ are the four-dimensional $SU(2)$ R-symmetry indices. Hence half of the supersymmetry is represented in the chiral representation and the other half in the anti-chiral representation. This non-chiral feature is generic for dimensional reduction of six-dimensional supersymmetric theories~\cite{HuangPeng, NonChiral} and stems from the fact that it is convenient to identify the four-dimensional $U(1)$ little group with the diagonal $U(1)$ subgroup of the six-dimensional $SU(2)\times SU(2)$.

The six-dimensional $\mathcal{N}$=(1,0) vector and tensor multiplets both reduce to four-dimensional $\mathcal{N}=2$ super-Yang-Mills. In terms of on-shell superfields, this is expressed as:
\eqa
\nonumber {\rm Vector}\;\;6D:\Psi_{\dot 1}(\eta^a)&\rightarrow&\;\;4D: \Psi^{4D}=\lambda^+_{2}+\zeta^1\phi_{12}+ \bar{\zeta}_2A^++\zeta^1 \bar{\zeta}_2\lambda^{+}_1\\
\nonumber \;\;\Psi_{\dot 2}(\eta^a)&\rightarrow&\quad\quad\quad\bar\Psi^{4D}=\bar\lambda^{1-}+\zeta^1A^-+\bar{\zeta}_2\bar{\phi}^{12}+\zeta^1 \bar{\zeta}_2\bar\lambda^{2-}
\eqae
The $SU(2)$ R-symmetry indices on the four-dimensional fields can be raised and lowered by the metric $\epsilon^{12}=\epsilon_{21}=1$. This leads to the following identification of states:
\eqa
\nonumber{\rm Vector} \;\;6D:\quad \left(\begin{array}{cc} G^1\,_{\dot 1} & G^1\,_{\dot 2} \\ G^2\,_{\dot 1} &  G^2\,_{\dot 2}\end{array}\right)\quad&\rightarrow&\;\nonumber4D:\quad\left(\begin{array}{cc}\phi_{12} & A^-  \\  A^+ & \bar\phi^{12}\end{array}\right)\\
 \quad\lambda_{\dot a},\;\tilde\lambda_{\dot a}\quad&\rightarrow&\quad\quad\quad\;\left(\begin{array}{c}\lambda^+_{2} \\ \bar\lambda^{1-}\end{array}\right),\;\left(\begin{array}{c} \lambda^{+}_1 \\ \bar\lambda^{2-}\end{array}\right)
\end{eqnarray}
For the $\mathcal{N}$=(1,0) tensor multiplet, one has:
\eqa
\nonumber{\rm Tensor}\;\;6D:\Psi^1(\eta^a)&\rightarrow&\;\;4D: \bar\Psi^{4D}=\bar\lambda^{1-}+\bar{\zeta}_2(\hat\phi^{21}+i\phi^{21})+\zeta^1A^{-}+\zeta^1 \bar{\zeta}_2\bar{\lambda}^{2-}\\
\Psi^2(\eta^a)&\rightarrow&\quad\quad\;\;\; \Psi^{4D}=\lambda^{+}_2+\zeta^1(\hat\phi_{12}-i\phi_{12})+\bar{\zeta}_2A^++\zeta^1 \bar{\zeta}_2\lambda_{1}^+\,.
\eqae
This leads to:
\eqa
\nonumber{\rm Tensor}\;\;6D:\quad \left(\begin{array}{cc} B^{11} & B^{12} \\ B^{12} &  B^{22}\end{array}\right)\quad&\rightarrow&\;\nonumber4D:\quad\left(\begin{array}{cc}A^- & -i\phi_{12}  \\  -i\phi^{12} & A^+\end{array}\right)\\
\nonumber \phi\quad&\rightarrow&\quad\quad\quad\;\hat\phi^{12}\\
 \quad \chi^a,\tilde\chi^a\quad&\rightarrow&\quad\quad\quad\;\left(\begin{array}{c}\bar\lambda^{1-} \\ \lambda^{+}_2\end{array}\right),\,\left(\begin{array}{c}\bar\lambda^{2-} \\ \lambda^{+}_1\end{array}\right)
\end{eqnarray}
The reduction of the scalar and the $\mathcal{N}$=(2,0) multiplets follows straightforwardly from the above. Under this reduction, four-dimensional positive (negative) helicity states come from the six-dimensional superfields with lower $SU(2)$ index 1 (2).

\section{General six-dimensional $\mathcal{N}=(1,0)$ three-point amplitudes}
Equipped with the convenient on-shell variables, the construction of three-point amplitudes becomes a simple question of what symmetries we wish to impose. In fact, requiring both $\alpha$-scaling and $b$-shift invariance, which amounts to six-dimensional Lorentz invariance, one can immediately rule out a three-point amplitude involving three self-dual tensors. To see this, note that such an amplitude will be required to carry six chiral $SU(2)$ little group indices. Requiring $\alpha$-scaling invariance immediately fixes the amplitude to be a product of three $u$s and three $w$s. It is then easy to see that such a polynomial cannot be made $b$-shift invariant. Thus, without attempting to non-abelianize the abelian gauge algebra or invoking strong coupling arguments, the three-point self-dual tensor interaction is ruled out simply by kinematics! 

An immediate generalization would be to allow for the self-dual tensors to interact with other fields. To limit our analysis, we focus on general six-dimensional three-point amplitudes that satisfy super-Poincar\'{e} invariance. Thus, we begin by looking for solutions to the supersymmetry constraints. 
\subsection{Supersymmetry in three-point kinematics}
From the previous discussion, we see that R-invariance fixes the Grassman degree of the $\mathcal{N}=(N,0)$ three-point amplitude to be $3N$. In four dimensions, half of the on-shell supersymmetry is enforced by a supermomentum-conserving delta function due to the multiplicative nature of the supercharges. Such a delta function in six dimensions would have degree $4N$ since each supermomentum carries an $SU^{*}(4)$ index. The mismatch of the Grassmann degrees shows that the three-point amplitude cannot be proportional to the supermomentum delta function\footnote{This statement is true for on-shell superspace that preserves little group invariance and breaks manifest R-symmetry.}. This contradiction can be resolved by noting that in three-point special kinematics, the supermomentum conservation constraints are not all linearly independent. Instead, the degree 4 supermomentum delta function gives rise to three linearly independent constraints. The same situation happens in four dimensions, where the $\overline{\rm MHV}$ three-point amplitude for supersymmetric Yang-Mills in the chiral representation is not proportional to a supermomentum delta function.

The $8N$ supersymmetry constraints are of two types, multiplicative and differential. We begin with the first:
\begin{equation}
Q^{mA+}=\sum_{i=1}^3 \lambda_i^{Aa}\eta^{m}_{ia}=0 \label{constraint1}
\end{equation}
Because these are fermionic multiplicative constraints, every independent component of (\ref{constraint1}) must be a factor of the amplitude in order for it to vanish after multiplication by the SUSY charge. From the previous paragraph we know that we are looking to derive three independent constraints from (\ref{constraint1}). Because the conditions corresponding to the different $Sp(2)$ subgroups of the R-symmetry are independent, it is sufficient to treat them in isolation. Thus, below we drop the superscripts $m$ from all the $Q$'s and $\eta$'s, remembering to include $N$ factors at the end of the day. 

The three linearly independent constraints can be found by decomposing the supercharge as follows (see \cite{Brandhuber:2010mm} for similar discussion):
\begin{equation}
Q^{A+}= \sum_{i=1}^3 \lambda_i^{Aa}\eta^{+b}_{i}\epsilon_{ab}
= \sum_{i=1}^3 \lambda_i^{Aa}\eta^{+b}_{i}u_{i[a}w_{ib]}\,
\end{equation}
Using eqs.~(\ref{Ueq}) and (\ref{Weq}), choose an independent basis for the bosonic spinor products. Selecting, for example,  $(\lambda_1^{Aa}u_{1a}, \lambda_2^{Aa}w_{2a},\lambda_3^{Aa}w_{3a})$, the decomposition then leads to 
\begin{equation}
Q^{A+}=-\lambda_1^{Aa}u_{1a} \left(\sum_{i=1}^3 \mathbf{w}_i\right) - \lambda_2^{Aa}w_{2a}(\mathbf{u}_1-\mathbf{u}_2)- \lambda_3^{Aa}w_{3a}(\mathbf{u}_1-\mathbf{u}_3)\,,
\end{equation}
where we have used the notation $\mathbf{u}_i\equiv u^a_i\eta_{ia}$ and $\mathbf{w}_i\equiv w^a_i\eta_{ia}$. Thus, the three independent constraints are simply:
\begin{eqnarray}
(\mathbf{u}_1-\mathbf{u}_2)=(\mathbf{u}_2-\mathbf{u}_3)&=&0 \label{FCons1}\\
\sum_{i=1}^3 \mathbf{w}_i&=&0
 \label{FCons2}
\end{eqnarray}
Since the above conditions are clearly sufficient to make $Q^{A+}$ vanish, the three-point amplitude must be proportional to 
\begin{eqnarray}
\nonumber\Delta(Q)&\equiv&\left[\mathbf{u}_1-\mathbf{u}_2) \right](\mathbf{u}_2-\mathbf{u}_3)\left(\sum_{i=1}^3{\mathbf{w}}_i\right)\\ 
&=&\left(\mathbf{u}_1\mathbf{u}_2+\mathbf{u}_2\mathbf{u}_3+\mathbf{u}_3\mathbf{u}_1\right)\left(\sum_{i=1}^3\mathbf{w}_i\right),
\label{deltaq}
\end{eqnarray}
where the notation $\Delta(Q)$ indicates that this constraint comes from the chiral supercharge. Indeed, in \cite{Bern:2010qa} the three-point amplitude of the $(1,1)$ sYM theory was found to precisely equal $\Delta(Q)\Delta(\tilde{Q})$.

As for the differential supersymmetry constraints, one easily verifies that
\begin{align}
Q^{A-} \Delta(Q) & = 
\sum_{i=1}^3 \lambda_i^{Aa}\frac{\partial}{\partial \eta_i^{ma+}} \left(\mathbf{u}_1\mathbf{u}_2+\mathbf{u}_2\mathbf{u}_3+\mathbf{u}_3\mathbf{u}_1\right)\left(\sum_{i=1}^3\mathbf{w}_i\right) \nonumber \\
& = \left(\lambda^{A}_1\cdot u_1(\mathbf{u}_2-\mathbf{u}_3)
   + \lambda^{A}_2\cdot u_2(\mathbf{u}_3-\mathbf{u}_1)
   + \lambda^{A}_3\cdot u_3(\mathbf{u}_1-\mathbf{u}_2)\right)
\left(\sum_{i=1}^3\mathbf{w}_i\right) \nonumber \\ 
& + \left(\mathbf{u}_1\mathbf{u}_2+\mathbf{u}_2\mathbf{u}_3+\mathbf{u}_3\mathbf{u}_1\right)\left(\sum_{i=1}^3\lambda_i^{Aa}w_{ia}\right),
\end{align}
which vanishes by eqs.~(\ref{Ueq}) and (\ref{Weq}). This confirms that the fermionic factor $\Delta(Q)$ is fully supersymmetric. In result, supersymmetry implies that the $\mathcal{N}=(N,0)$ three-point amplitude can be written as
\eq
\mathcal{A}_3=\delta^6(P)\left[ \left(\mathbf{u}_1\mathbf{u}_2+\mathbf{u}_2\mathbf{u}_3+\mathbf{u}_3\mathbf{u}_1\right)\left(\sum_{i=1}^3\mathbf{w}_i\right)\right]^Nf(u,w)=\delta^6(P) [\Delta(Q)]^Nf(u,w)\,.
\label{ansatz}
\eqe
The function $f(u,w)$ may carry free little group indices, depending on the on-shell multiplets. 
\subsection{$\mathcal{N}=(1,0)$ three-point amplitudes}
As discussed previously, supersymmetry is preserved in the presence of the fermionic factor $\Delta(Q)$. Hence, in constructing general supersymmetric amplitudes with $\mathcal{N}=(N,0)$ supersymmetry, it suffices to identify appropriate functions $f(u,w)$. They must scale as $\alpha^{-1}$ under scaling (\ref{alpha}), because $\Delta(Q)$ scales as $\alpha$, and they must be invariant under the $b$-shift.

We first consider $\mathcal{N}=(1,0)$ with only vector, scalar, and tensor multiplets involved. The $\alpha$-scaling requirement leads to the following possibilities\footnote{One could consider higher degrees of ($u,w,\tilde{u},\tilde{w}$), but this would imply the inclusion of higher spin multiplets.}:
\eq
f(u,w)=(\tilde{u},w,\tilde{u}u\tilde{u},\tilde{u}uw,\tilde{u}\tilde{u}\tilde{w},\tilde{u}\tilde{w}w,wuw,w\tilde{u}\tilde{w},w\tilde{w}w)
\eqe
The $b$-shift invariance restricts these down to:
\begin{itemize}
  \item $f(u,w)=\tilde{u}_1$: this describes the $\mathcal{N}=(1,0)$ super-Yang-Mills coupling to two scalar multiplets,
  \item $f(u,w)=u_1 \tilde{u}_2\tilde{u}_3$: this describes a tensor multiplet coupled to two vectors,
  \item $f(u,w)=\tilde{u}_1\tilde{u}_2\tilde{w}_3+\tilde{u}_2\tilde{u}_3\tilde{w}_1+\tilde{u}_3\tilde{u}_1\tilde{w}_2$: this describes the self-interaction of $\mathcal{N}=(1,0)$ super-Yang-Mills.
  \end{itemize}
Notice the absence of a tensor-tensor-vector interaction, which is present in the recently proposed construction~\cite{Samtleben:2011fj}. We will discuss this apparent contradiction in section~\ref{WhatsWrong}. Note that these amplitudes are not dilation invariant, hence super-Poincar\'{e} invariance alone rules out any conformal three-point amplitude.

We could attempt the same procedure for $\mathcal{N}=(2,0)$ and look for amplitudes that involve only the self-dual tensor multiplet. With the on-shell superfield being a scalar, i.e. carrying no spinor indices, one can only multiply the fermionic part by a Lorentz and little group scalar. Thus, the three-point $\mathcal{N}=(2,0)$ supersymmetric amplitude can only take one form:
\eqa
\mathcal{N}=(2,0):\quad\delta^6(P)\mathcal{A}_3=\left[ \Delta(Q) \right]^2
\eqae
However, this is not $\alpha$-scaling invariant. Therefore, three-point amplitudes with pure  $\mathcal{N}=(2,0)$ tensor multiplets are ruled out simply by super-Poincar\'{e} invariance.

We now turn to a study of the symmetries of the above interactions and their reduction to four dimensions. 

\subsubsection{The $\mathcal{N}=(1,0)$ vector-scalar-scalar interaction}
To study the dimensional reduction of six-dimensional massless amplitudes, we will start with the vector-scalar interactions, which should trivially reduce to the four-dimensional Yang-Mills-scalar three-point amplitude. Consider the  scalar Yang-Mills amplitude
\eqa
\nonumber\mathcal{A}_3(\Phi(i)\bar\Phi(j)\Psi_{\dot{a}}(k))&=&\delta^6(P)\Delta(Q)(\tilde{u}_k)_{\dot{a}},
\eqae
where the $k^{\rm th}$ leg represents the position of the vector multiplet. From the map between six- and four-dimensional states given in section \ref{Map} we see that the negative helicity vector field in four dimensions corresponds to $G^1\,_{\dot{2}}$. Thus, a typical term that should map to the four-dimensional $A(\phi\phi-)$ amplitude will come from the $(\eta_2)_1(\eta_2)_2(\eta_3)_1$ component of $\mathcal{A}_3(\Phi(1)\bar\Phi(2)\Psi_{\dot{2}}(3))$. Plugging in the specific four-dimensional solution to the $u,\tilde{u}$ variables given in eq.~(\ref{expl6}) gives
\eq
\mathcal{A}_3(\Phi(1)\bar\Phi(2)\Psi_{\dot{2}}(3))|_{(\eta_2)_1(\eta_2)_2(\eta_3)_1}=(u_3)^1(\tilde{u}_3)_{\dot{2}}=\frac{\langle23\rangle\langle31\rangle}{\langle12\rangle},
\eqe
which is precisely the MHV amplitude $A(\phi\phi-)$. Note that while the solutions in eq.~(\ref{expl6}) are defined up to $\alpha$-scaling, since the amplitude is $\alpha$-scaling invariant, any other explicit solution for the $u,\tilde{u}$ variables will give the same result.

\subsubsection{The $\mathcal{N}=(1,0)$ vector-vector-tensor interaction\label{proto}}
We first check the symmetry of the amplitude when one exchanges two vector fields. Since the definition of the $u,\tilde{u}$ in eq.~(\ref{not}) picks up an additional minus sign if one exchanges any two legs, this means that $u_i$ and $\tilde u_i$ uniformly picks up a phase $i$ under the exchange of two legs, and a $-i$ for $w$ and $\tilde{w}$. Thus, we have:
\eq
\left(\mathbf{u}_1\mathbf{u}_2+\mathbf{u}_2\mathbf{u}_3+\mathbf{u}_3\mathbf{u}_1\right)\left(\sum_{i=1}^3\mathbf{w}_i\right)\xrightarrow{1\leftrightarrow 2}-i\left(\mathbf{u}_1\mathbf{u}_2+\mathbf{u}_2\mathbf{u}_3+\mathbf{u}_3\mathbf{u}_1\right)\left(\sum_{i=1}^3\mathbf{w}_i\right)
\eqe
Combing the phase factors coming from the bosonic part $\tilde{u}_{1\dot{a}}\tilde{u}_{2\dot{b}}u_{3c}$ as well as the additional minus sign coming from the exchange of fermionic superfields, we see that the amplitude has a minus sign under the exchange of the two vector legs. Consequently, the coupling constant of the three-point vertex must be symmetric under the exchange of two vector indices\footnote{As an exercise, one can apply the same argument to super-Yang-Mills and conclude that $f_{ijk}=-f_{jki}$.}:
\eq
\mathcal{A}_3=\langle\Psi^i_{\dot{a}}(1)\Psi^j_{\dot{b}}(2)\Psi^{ka}(3)\rangle f_{ijk},\quad\quad f_{ijk}=f_{jki}
\eqe 
Let us now look at the $\langle B(1)A(2)A(3)\rangle$ amplitude, which is proportional to $(\eta_1)_d(\eta_2)_b(\eta_3)_c$ with symmetrized $SU(2)$ indices on the first leg. It is: 
\eqa
\nonumber&&\langle B(1)G(2)G(3)\rangle^{(da)bc}\,_{\dot{a}\dot{b}}\\
\nonumber&=&\delta^6(P)\left[(u_2)^b(w_3)^c(u_1)^{(d}+(u_2)^b(u_3)^c(w_1)^{(d}+(u_3)^c(w_2)^b(u_1)^{(d}\right](u_1)^{a)}(\tilde{u}_2)_{\dot{a}}(\tilde{u}_3)_{\dot{b}}\\
\label{BAA1}
\eqae
Again, eq.~(\ref{BAA1}) is symmetric under the exchange of $2\leftrightarrow3$. Being six-dimensional Lorentz covariant and mass dimension two\footnote{Here the mass dimension of $\delta^6(P)$ is not counted.}, the symmetric form of the amplitude fixes the tensor-vector-vector interaction term in the action to be 
\eq
\mathcal{L}_3\sim\epsilon^{\mu\nu\rho\sigma\tau\upsilon}B_{\mu\nu}F_{\rho\sigma}\bar{F}_{\tau\upsilon},
\label{SiegelTerm}
\eqe
where we have allowed the two vector multiplets to be distinct. This term appeared in the interacting self-dual tensor-vector action written down by Siegel in ref.~\cite{Siegel:1983es}, where the gauge symmetry is purely abelian. It also appears in explicit actions of $\mathcal{N}=(1,0)$ supergravity coupled to various $\mathcal{N}=(1,0)$ matter multiplets~\cite{Nishino:1984gk,Romans:1986er} as well as the recent  $\mathcal{N}=(1,0)$ action proposed in~\cite{Samtleben:2011fj}.

Now let us look at the four-dimensional massless reduction of eq.~(\ref{BAA1}). From the map between six and four-dimensional states, we see that the four-dimensional MHV amplitude $A(--+)$ comes from $\langle B(1)G(2)G(3)\rangle^{1112}\,_{\dot{2}\dot{1}}$ while $A(+--)$ comes from $\langle B(1)G(2)G(3)\rangle^{2211}\,_{\dot{2}\dot{2}}$. From the explicit solutions in eqs.~(\ref{expl6}) and (\ref{explw}) it is easy to see that the amplitude in eq.~(\ref{BAA1}) vanishes for these configurations. A similar analysis shows that eq.~(\ref{BAA1}) vanishes for the configuration that descends to $A(---)$ as well. In fact, one can see that the only non-vanishing four-dimensional massless descendant of the above amplitude involves scalars. As an example, consider the component that descends to the four-dimensional amplitude $(\phi A^- A^-)$: 
\eqa
\nonumber&&\langle B(1)G(2)G(3)\rangle^{(12)11}\,_{\dot{2}\dot{2}}\\
\nonumber&=&\left[(u_2)^1(w_3)^1(u_1)^{(1}+(u_2)^1(u_3)^1(w_1)^{(1}+(u_3)^1(w_2)^1(u_1)^{(1}\right](u_1)^{2)}(\tilde{u}_2)_{\dot{2}}(\tilde{u}_3)_{\dot{2}}\\
&=&\langle23\rangle^2
\eqae
The component that reduces to $(\phi A^+ A^+)$, which is $\langle B(1)G(2)G(3)\rangle^{(12)22}\,_{\dot{1}\dot{1}}$, can be computed using eqs.~(\ref{expl62}) and (\ref{explw2}) and gives $-[23]^2$. Thus, one sees that after dimensional reduction these amplitudes correspond to the following term in the effective action:
\eq
4D: \quad\langle23\rangle^2-[23]^2\rightarrow\phi(1)(f^{\alpha\beta}(2)f_{\alpha\beta}(3)-\tilde{f}^{\dot{\alpha}\dot{\beta}}(2)\tilde{f}_{\dot{\alpha}\dot{\beta}}(3))=\phi F\wedge F
\eqe
Here $f_{\alpha\beta}$ and $\tilde{f}_{\alpha\beta}$ are the self-dual and anti-self-dual fields strengths, respectively. This is indeed the expected result of a dimensional reduction of the six-dimensional term given in eq.~(\ref{SiegelTerm}). 

Thus, we conclude that our amplitude is simply an $\mathcal{N}=(1,0)$ supersymmetrization of the bosonic interaction in eq.~(\ref{SiegelTerm}).

\subsection{The lack of a vector-tensor-tensor interaction \label{WhatsWrong}}
In the above analysis, a conspicuous problem is the absence of tensor-tensor-vector interactions, which appear in the action proposed in~\cite{Samtleben:2011fj}. Explicitly, such interaction arises from $H^i_{\mu\nu\rho}H^{i\mu\nu\rho}$, where the three-form field strength  $H^i_{\mu\nu\rho}$ is given by
\eq
H^i_{\mu\nu\rho}=\partial_{[\mu} B^i_{\nu\rho]}+f^{ijk}A^{k}_{[\mu} B^{j}_{\nu\rho]}+\cdot\cdot\cdot\, ,
\eqe 
with $\cdot\cdot\cdot$ including a cubic term in vectors and a three-form tensor. The vector-tensor-tensor interaction is given by
\eq
f^{ijk}\partial_{[\mu} B^i_{\nu\rho]}A^{k[\mu} B^{j\nu\rho]}
\label{threepoint}
\eqe 
and the linearized gauge transformations of the vector and tensor field are:
\eqa
\nonumber\delta A_\mu&=&\partial_\mu \Lambda+\Lambda_\mu+\cdot\cdot\cdot\\
\delta B_{\mu\nu}&=&\partial_{[\mu} \Lambda_{\nu]}+\cdot\cdot\cdot\,
\label{gauge}
\eqae
We now comment on the reason why eq.(\ref{threepoint}) cannot give a consistent amplitude. A hint can be found in the form of polarization vectors and tensors, written in terms of on-shell variables: 
\eqa
\nonumber\tilde{\sigma}^{\nu AB}(\epsilon_{\nu})_{ a\dot b}=\epsilon^{AB}_{a\dot b}(\mu,p_i)&=&\frac{\,^{[A}|i_{a}\rangle[i_{\dot{b}}|_C\mu^{B]C}}{s_{i,\mu}}\\
\tilde{\sigma}^{\nu AB}\tilde{\sigma}^{\rho}_{BC}(\tilde\epsilon_{\nu\rho})_{ab}=\tilde\epsilon^{A}\,_{C,ab}(\mu,p_i)&=&\frac{\,^{A}|i_{(a}\rangle\langle i_{b)}|^B\mu_{BC}}{s_{i,\mu}}\,,
\label{pol}
\eqae
where $\mu$ is a null reference vector satisfying $2p_i\cdot\mu=s_{i\mu}\neq0$. These polarization vectors and tensors manifestly satisfy the Lorentz gauge condition, $p_i\cdot\epsilon_i=p_{i\mu}\tilde\epsilon_i^{\mu\nu}=0$. It can be straightforwardly seen, following the arguments in~\cite{cheung0}, that a change of reference vector induces the following gauge transformation:
\eqa
\nonumber\epsilon^{'AB}_{a\dot b}(\mu',p_i)&=&\epsilon^{AB}_{a\dot b}(\mu,p_i)+p_i^{AB}\Omega_{a\dot b}\\
\tilde\epsilon^{A}\,_{C,ab}(\mu',p_i)&=&\tilde\epsilon^{A}\,_{C,ab}(\mu,p_i)+p_i^{AB}\Omega_{BC(ab)}-p_{iCB}\Omega^{BA}_{(ab)}\,
\eqae 
Note that the gauge transformation of the vector field, $\epsilon^{AB}_{a\dot b}$, lacks the $\Lambda_\mu$ part in eq.~(\ref{gauge}). The reason is simple: the $\Lambda_\mu$ gauge symmetry is not a symmetry of the usual free Maxwell action. Since the polarization vectors and tensors represent asymptotic states which are solutions to the free action, it only has the gauge symmetries of the free action.

As a consequence, if we insert the polarization tensors into the three-point interaction term given in eq.~(\ref{threepoint}), it will not be a gauge invariant quantity, because the cubic term in the action is gauge invariant at the linear level only if the vector field transforms as in eq.~(\ref{gauge}). One can confirm this by explicit substitution: while the reference vector for the polarization vector drops out of the final result, the reference vector for the polarization tensor will not, indicating the non-gauge invariance with respect to the $\Lambda_\mu$ transformation.  

From this one concludes that the tensor-tensor-vector amplitude derived from eq.(\ref{threepoint}) cannot be made gauge invariant. This seems to contradict the work of~\cite{Samtleben:2011fj}, where gauge invariance holds. One can pinpoint the issue by looking at the quadratic part of the action that contains the vector field strength. In~\cite{Samtleben:2011fj} in the broken phase, it is given by $F^2$ with
\eq
F_{\mu\nu}=\partial_{[\mu}A^r_{\nu]}+h^r_IB^I_{\mu\nu}-f_{st}\,^rA_\mu^s A_\nu^t,\;\
\eqe
where $h^r_I$ is some group-dependent coupling constant. In the free limit, indeed the $\Lambda_\mu$ gauge symmetry is not a symmetry of the action, $F^2$. However, once the interaction is turned on, the quadratic part of the action becomes 
\eq
(\partial_{[\mu}A^r_{\nu]}+h^r_IB^I_{\mu\nu})(\partial^{[\mu}A^{\nu]r}+h^r_JB^{J\mu\nu}).
\eqe
This leads to a mixing of the fields, and quantization becomes difficult as the propagator itself will be coupling constant-dependent. Furthermore, as the usual perturbation theory requires one to expand around the quadratic term of the action, here the asymptotic states defined by the free theory cannot be used for computing the perturbative expansion.  

In conclusion, we see that the difficulty in obtaining amplitudes from these theories lies in the fact that the eigenstates of the kinetic terms are not those of the free theory, and the definition of asymptotic states becomes problematic.
\section{$\mathcal{N}=(2,0)$ Amplitudes with higher spins}
It is very curious that we could not obtain an amplitude with three $\mathcal{N}=(2,0)$ tensor multiplets, even when only super-Poincar\'{e} invariance was required. Motivated by the assumption that the interactions of separated M5 branes can be mediated via self-dual strings in their world volume, we generalize the analysis to allow for higher spin $\mathcal{N}=(2,0)$ multiplets. This allows us to introduce additional $u,\tilde{u},w,\tilde{w}$ variables to $f(u,w)$. We immediately have a large class of three-point amplitudes. Considering only amplitudes that do not involve anti-self-dual fields, we have: 
\eqa
\nonumber(1)\quad\mathcal{A}_3^{\mathcal{N}=(2,0)}&=&\delta^6(P)\left[ \Delta(Q) \right]^2\tilde{u}_{i\dot{a}}\tilde{u}_{j\dot{b}}\\
\nonumber(2)\quad\mathcal{A}_3^{\mathcal{N}=(2,0)}&=&\delta^6(P)\left[ \Delta(Q) \right]^2\left(u_{1a}u_{2b}w_{3c}+u_{1a}w_{2b}u_{3c}+w_{1a}u_{2b}u_{3c}\right)\tilde{u}_{i\dot{a}}\tilde{u}_{j\dot{b}}\tilde{u}_{k\dot{c}}\\
(3)\quad\mathcal{A}_3^{\mathcal{N}=(2,0)}&=&\delta^6(P)\left[ \Delta(Q) \right]^2\tilde{u}_{i\dot{a}}\tilde{u}_{i\dot{b}}
\label{cookie}
\eqae
We discuss the details of the multiplets involved in the interaction:
\begin{itemize}
  \item $(1)\quad\mathcal{A}_3^{\mathcal{N}=(2,0)}=\delta^6(P)\left[ \Delta(Q) \right]^2\tilde{u}_{i\dot{a}}\tilde{u}_{j\dot{b}}$: This amplitude involves one $\mathcal{N}=(2,0)$ tensor multiplet along with two fields of the following multiplet: 
  \eqa
\nonumber  \Psi^{\dot{a}}(\eta,\hat{\eta})&=&\psi^{\dot{a}}+\eta^{a}G_{a}\,^{\dot{a}}+\hat{\eta}^{a}\bar{F}_{a}\,^{\dot{a}}+\eta^2\chi^{\dot{a}}+(\eta^a\hat{\eta}_a)\lambda^{\dot{a}}+\hat{\eta}^2\hat{\chi}^{\dot{a}}+\eta^{(a}\hat{\eta}^{b)}S_{(ab)}\,^{\dot{a}}\\
&&\eta^2\hat{\eta}^aF_a\,^{\dot{a}}+\hat{\eta}^2\eta^a\bar{G}_a\,^{\dot{a}}+\eta^{2}\hat{\eta}^{2}\bar{\psi}^{\dot{a}}
\label{SelfDualSpin}
  \eqae 
  We see that this multiplet contains 4 vectors as the bosonic fields, while the fermions involve a spin-$3/2$ fermion $S_{(ab)}\,^{\dot{a}}$. 
  \item $(2)\quad\mathcal{A}_3^{\mathcal{N}=(2,0)}=\left[ \Delta(Q) \right]^2\left(u_{1a}u_{2b}w_{3c}+u_{1a}w_{2b}u_{3c}+w_{1a}u_{2b}u_{3c}\right)\tilde{u}_{i\dot{a}}\tilde{u}_{j\dot{b}}\tilde{u}_{k\dot{c}}$: This amplitude involves fields in the following multiplet:
  \eqa
  \nonumber A^{a\dot{a}}(\eta,\hat\eta)&=&G_1^{a\dot{a}}+\eta_b\left(\epsilon^{ba}\chi^{\dot{a}}+S^{(ba)\dot{a}}\right)+\hat{\eta}_b\left(\epsilon^{ba}\hat{\chi}^{\dot{a}}+\hat{S}^{(ba)\dot{a}}\right)\\
  \nonumber &&+\eta^2G_2^{a\dot{a}}+\hat{\eta}^2G_3^{a\dot{a}}+\eta^b\hat{\eta}_bG_4^{a\dot{a}}+\eta_{(b}\hat{\eta}_{c)}\left[D^{(bca)\dot{a}}+\epsilon^{a(b}G_5^{c)\dot{a}}\right]\\
\nonumber &&+\eta^2\hat{\eta}_b\left(\epsilon^{ba}\chi^{'\dot{a}}+S^{'(ba)\dot{a}}\right)+\hat\eta^2\eta_b\left(\epsilon^{ba}\hat\chi^{'\dot{a}}+\hat{S}^{'(ba)\dot{a}}\right)+\eta^2\hat\eta^2G_6^{a\dot{a}}
  \eqae 
 This multiplet includes 6 vectors and one self-dual rank 3 tensor that also appeared in the $\mathcal{N}=(3,0)$ multiplet. In the next section, we will show that this amplitude is actually a supersymmetry truncation of an $\mathcal{N}=(3,0)$ three-point amplitude involving three self-dual rank 3 tensor multiplets.
 
   \item $(3)\quad\mathcal{A}_3^{\mathcal{N}=(2,0)}=\left[ \Delta(Q) \right]^2\tilde{u}_{i\dot{a}}\tilde{u}_{i\dot{b}}$: This amplitude involves two $\mathcal{N}=(2,0)$ tensor multiplets along with one $\mathcal{N}=(2,0)$ supergravity multiplet:
  \eqa
  \nonumber\Phi^{(\dot{a}\dot{b})}(\eta,\hat\eta)&=&B^{(\dot{a}\dot{b})}+\eta^a\psi_a\,^{(\dot{a}\dot{b})}+\hat{\eta}^a\hat{\psi}_a\,^{(\dot{a}\dot{b})}+\eta^2B^{'(\dot{a}\dot{b})}+\hat{\eta}^2B^{''(\dot{a}\dot{b})}+\eta^a\hat{\eta}_aB^{'''(\dot{a}\dot{b})}\\
  \nonumber&& +\eta_{(a}\hat{\eta}_{b)}A_{(ab)}\,^{(\dot{a}\dot{b})}+\eta^2\hat{\eta}^a\hat{\bar{\psi}}_a\,^{(\dot{a}\dot{b})}+\hat{\eta}^2\eta^a\bar{\psi}_a\,^{(\dot{a}\dot{b})}+\eta^2\hat{\eta}^2B^{''''(\dot{a}\dot{b})}
  \eqae
The bosonic fields include 5 anti-self-dual two-forms, a graviton that transforms as a $(\mathbf{3},\mathbf{3})$ under the little group and 4 gravitinos. Again note that the gravitinos $\psi_a\,^{(\dot{a}\dot{b})}$ have the opposite chirality to the self-dual spinor $S^{(ba)\dot{a}}$.
\end{itemize}

The above amplitudes show that at three points, a two-tensor multiplet interaction is only possible through a gravitational coupling. This result is reminiscent of the $\mathcal{N}=(1,0)$ story, where for non-gravitational amplitudes, the self-dual tensor multiplet can only interact through two vector multiplets. 
\subsection{$\mathcal{N}=(2,0)$ Self-dual tensor-spin-3/2 multiplet coupling}
With the goal of finding a non-gravitational coupling involving self-dual tensor multiplets, we focus on case (1) in eq.~(\ref{cookie}). The two-vector, one self-dual tensor component amplitude $\langle B(1)A(2)A(3)\rangle$ corresponds to the  $(\eta_1)_{(a}(\hat{\eta}_1)_{b)}(\hat{\eta}_2^2)(\eta_2)_c(\eta_3)_d$ component of this case:
\eqa
\nonumber&&\langle B(1)G(2)G(3)\rangle^{(ab)cd}\,_{\dot{c}\dot{d}}=\\
\nonumber&&\delta^6(P)\left[(u_2)^c(w_3)^d(u_1)^{(a}+(u_2)^c(u_3)^d(w_1)^{(a}+(u_3)^d(w_2)^c(u_1)^{(a}\right](u_1)^{b)}(\tilde{u}_{2})_{\dot{c}}(\tilde{u}_{3})_{\dot{d}}\,.\\
\eqae
Following the discussion in section \ref{proto}, one sees that the above amplitude is symmetric under the exchange of legs 2 and 3. This again leads to the following possible cubic term in the action:
\eq
\mathcal{L}_3\sim \epsilon^{\mu\nu\rho\sigma\tau\upsilon}B_{\mu\nu}F_{\rho\sigma}\bar{F}_{\tau\upsilon}\,
\label{BAA2}
\eqe
We test this proposal by looking at the four-dimensional reduction, which by eq.~(\ref{BAA2}) would lead to vanishing results for $A(---)$, $A(--+)$, and $A(+--)$. One can explicitly check that the corresponding configurations $\langle B(1)G(2)G(3)\rangle^{(11)11}\,_{\dot{2}\dot{2}}$, $\langle B(1)G(2)G(3)\rangle^{(11)12}\,_{\dot{2}\dot{1}}$, and $\langle B(1)G(2)G(3)\rangle^{(22)11}\,_{\dot{2}\dot{2}}$ indeed vanish. Checking the component that reduces to $A(\phi--)$ one has:
\eqa
\nonumber&&\langle B(1)G(2)G(3)\rangle^{(12)11}\,_{\dot{2}\dot{2}}=\\
\nonumber&&\delta^4(P)\left[(u_2)^1(w_3)^1(u_1)^{(1}+(u_2)^1(u_3)^1(w_1)^{(1}+(u_3)^1(w_2)^1(u_1)^{(1}\right](u_1)^{2)}(\tilde{u}_{2})_{\dot{c}}(\tilde{u}_{3})_{\dot{2}}\\
&&=\langle23\rangle^2
\eqae  
Combing with the component that reduces to $A(\phi++)$, one confirms that the above is a reduction of eq.~(\ref{BAA2}).

In the $\mathcal{N}=(1,0)$ analysis we commented that the bosonic coupling of the form in eq.~(\ref{BAA2}) appears in minimal supergravity theories. Thus, one might wonder if the above amplitude can be obtained from truncating a supergravity theory. Using KLT relations~\cite{Kawai:1985xq}, the three-point amplitude of the maximal $\mathcal{N}=(2,2)$ supergravity is given by 
\eq
\mathcal{A}^{\mathcal{N}=(2,2)}_3=\Delta(Q)\Delta(\hat{Q})\Delta(\tilde{Q})\Delta(\hat{\tilde{Q}}),
\eqe
where we have used $\tilde{Q}$ and $\hat{\tilde{Q}}$ to represent the anti-chiral supercharges of $\mathcal{N}=(2,2)$. Picking out the terms in the amplitude proportional to $(\tilde{\eta}_{2\dot{b}}\tilde{\eta}_{2}^{\dot{b}})(\hat{\tilde{\eta}}_{3\dot{a}}\hat{\tilde{\eta}}_{3}^{\dot{a}})$, one finds:
\eq
\int d^2\tilde{\eta}_2d^2\hat{\tilde{\eta}}_3\mathcal{A}^{\mathcal{N}=(2,2)}_3=\Delta(Q)\Delta(\hat{Q})(\tilde{\mathbf{u}}_3-\tilde{\mathbf{u}}_1)(\hat{\tilde{\mathbf{u}}}_1-\hat{\tilde{\mathbf{u}}}_2)
\eqe 
Picking the $(\hat{\tilde{\eta}}_1)_{\dot{a}}(\tilde{\eta}_3)_{\dot{b}}$ component, one recovers case (1) in eq.~(\ref{cookie}) with ($i$=1, $j$=3). Thus, this amplitude can indeed be embedded in the $\mathcal{N}=(2,2)$ supergravity amplitude.

One can ask, however, if the amplitude can be consistent by itself. Indeed, while $\mathcal{N}\leq2$ sYM amplitudes can be imbedded in $\mathcal{N}=4$ sYM amplitudes, these amplitudes are consistent as a subset. To test this, we can use the three-point amplitude to reconstruct the four-point amplitude via BCFW~\cite{Britto:2004ap,Britto:2005fq} recursion relations. When we shift legs 1 and 2, which stand for self-dual spinor multiplets, we obtain a four-point amplitude that factorizes correctly in the $t$-channel, as shown in diagram (a) of fig. \ref{4ptbcfw}. Following a similar computation in~\cite{cheung0,Dennen:2009vk}, we find: 
\begin{figure}
\begin{center}
\includegraphics[scale=0.9]{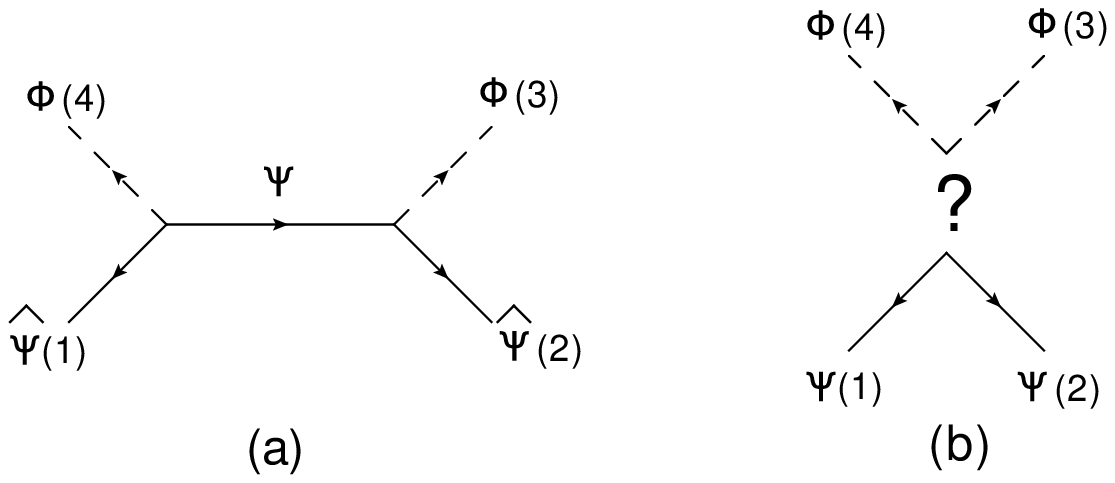}
\caption{ }
\label{4ptbcfw}
\end{center}
\end{figure}
\eq
\mathcal{A}_4(\Psi(1)\Psi(2)\Phi(3)\Phi(4))^{\dot{a}\dot{b}}=\delta^4(Q_{full})\delta^4(\hat{Q}_{full})\frac{[4^{\dot{a}}|p_1|3^{\dot{b}}]}{st}
\eqe
This result shows that the amplitude must also have a non-vanishing residue on the $s$-channel pole. This would predict a two self-dual tensor three-point amplitude mediated by an unknown multiplet, as shown in diagram (b) of fig.~\ref{4ptbcfw}. But from the discussion in the beginning of this section, we know that a two self-dual tensor three-point amplitude with $\mathcal{N}=(2,0)$ supersymmetry necessarily involves a gravitational multiplet. Thus, we see that the self-dual tensor-spinor amplitude has to be part of a supergravity amplitude. The spin-3/2 particle turns out to be the gravitino for the anti-chiral supersymmetry, which was not manifest.  

\subsection{General features of tensor coupling}
From the discussion of $\mathcal{N}=(1,0)$ and $\mathcal{N}=(2,0)$ three-point amplitudes that involve self-dual tensor multiplets we can extract some interesting, common features:
\begin{itemize}
  \item \textit{Three-point amplitudes that involve two tensor multiplets need gravity}: Both for $\mathcal{N}=(1,0)$ and $\mathcal{N}=(2,0)$ supersymmetry, three-point amplitudes that involve tensor multiplets require two of the fields to belong to another multiplet. The only two-tensor multiplet interaction is through a gravitational multiplet.
  
  \item \textit{Supersymmetry introduces anti-chiral fermion couplings}: For both cases, the multiplets that couple to the self-dual tensor multiplet contain anti-chiral fermions. This is a consequence of the amplitude being a supersymmetrization of the $B\wedge F\wedge F$ term, where the vector multiplet includes anti-chiral fermions.
  \item \textit{The universal appearance of $BF\bar{F}$}: It appears that the self-dual tensor contributes to the three-point amplitude by coupling with vector fields through: 
  \eq
\mathcal{L}_3 \sim \epsilon_{\mu\nu\rho\sigma\tau\upsilon}B^{\mu\nu}F^{\rho\sigma}\bar{F}^{\tau\upsilon}
\label{one}
  \eqe
  One can also check that the scalar from the tensor multiplet interacts with the vector through:
  \eq
\mathcal{L}_3\sim  \phi F^{\mu\nu}\bar{F}_{\mu\nu}\,
\label{two}
  \eqe
  This is simply a consequence of supersymmetry. Both these terms appear in  $\mathcal{N}=(1,0)$ and  $\mathcal{N}=(2,2)$ supergravity actions.
 \end{itemize}
Thus, super-Poincar\'{e} invariance alone ensures that all non-trivial three-point amplitudes involving tensor multiplets belong to supergravity systems.  As previously discussed, the non-gravitational $\mathcal{N}=(1,0)$ theory proposed in~\cite{Samtleben:2011fj}, while also containing interaction terms in eq.(\ref{one}) and eq.(\ref{two}), does not fit our analysis due to the difficulty of defining asymptotic states. 
\section{$\mathcal{N}=(3,0)$ self-dual three-point amplitudes}
Motivated by the analysis of interactions with higher spins, we study general supersymmetric interactions that only involve higher spins that are self-dual. We first consider $\mathcal{N}=(3,0)$. One immediately sees that there is a unique solution:
\eqa
\mathcal{N}=(3,0):\quad \mathcal{A}_3=\delta^6(P)\left[ \Delta(Q) \right]^3\tilde{u}_{1\dot{\alpha}}\tilde{u}_{2\dot{\beta}}\tilde{u}_{3\dot{\gamma}}
\label{303pt}
\eqae
Note that this amplitude carries anti-chiral $SU(2)$ indices, which indicates that the participating SUSY multiplet is the $\mathcal{N}=(3,0)$ multiplet discussed in section~\ref{multiplets}. Thus, amplitude~(\ref{303pt}) describes interactions of three $\mathcal{N}=(3,0)$ self-dual rank 3 tensor multiplets.

We can continue to $\mathcal{N}=(4,0)$ and further. Now, because the fermionic part has $\alpha$-weight four and higher, we see that non-self dual, higher spin fields are necessarily involved. This follows from the fact that canceling the $\alpha$-weight requires at least four $\tilde{u}$ variables, so in a three-point amplitude one will be forced to consider
\eq
f(u,w)\sim\tilde{u}_i^{\dot{a}}\tilde{u}_i^{\dot{b}},
\eqe  
which is symmetric in $\dot{a},\dot{b}$. Hence, the participating multiplet will involve anti-self-dual fields or gravitons. This leads to the conclusion that for pure self-dual interactions, at three points one can only have manifest supersymmetry up to $\mathcal{N}=(3,0)$.

We now analyze the form of this interaction in terms of component fields. Following similar arguments as for the vector-vector-tensor interaction, we see that 
\eq
\mathcal{A}_3^{\mathcal{N}=(3,0)}=\delta^6(P)\left[\Delta(Q)\right]^3\tilde{u}_{1\dot{a}}\tilde{u}_{2\dot{b}}\tilde{u}_{3\dot{c}}\xrightarrow{\quad1\leftrightarrow 2\quad}\delta^6(P)\left[\Delta(Q)\right]^3\tilde{u}_{1\dot{a}}\tilde{u}_{2\dot{b}}\tilde{u}_{3\dot{c}}\,.
\eqe
Again, taking into account the exchange of fermionic superfields, we conclude that the amplitude is totally antisymmetric with exchange of any two legs. To get a feeling of what kinds of interaction are included, let us consider the three-vector interaction. A typical term comes from $(\eta_1^m)^2(\eta^n_1)_{a}(\eta_2^n)^2(\eta^p_2)_{b}(\eta_3^p)^2(\eta^m_3)_{c}$, which gives:
\eq
\langle G(1)G(2)G(3)\rangle^{abc}\,_{\dot{a}\dot{b}\dot{c}}=\delta^6(P)(u_1)^a(u_2)^b(u_3)^c(\tilde{u}_1)_{\dot{a}}(\tilde{u}_2)_{\dot{b}}(\tilde{u}_3)_{\dot{c}}
\eqe
Being mass dimension three and totally anti-symmetric, it leads to the following three-vector interaction:
\eq
\mathcal{L}_3\sim(F_{1})^\mu\,_\nu(F_{2})^\nu\,_\rho(F_{1})^\rho\,_\mu
\label{F3}
\eqe
We can again test this by going to four dimensions. Using the explicit solutions from eq.~(\ref{expl6}), the component which reduces to $A(---)$ is:
\eq
\langle G(1)G(2)G(3)\rangle^{111}\,_{\dot{2}\dot{2}\dot{2}}=\langle12\rangle\langle23\rangle\langle31\rangle
\eqe
This corresponds to the four-dimensional amplitude
\eq
4D: \quad f_{\alpha}\,^{\beta}(1)f_{\beta}\,^{\gamma}(2)f_{\gamma}\,^{\alpha}(3),
\eqe
which is the self-dual part of eq.~(\ref{F3}) in four dimensions.

\section{Five-dimensional KK mode interactions\label{mass}}
The lack of non-gravitational self-dual tensor amplitudes may be attributed to the fact that the degrees of freedom are incorrect. While the self-dual tensor multiplet comprises the Goldstone bosons and fermions arising in the presence of M5 branes, their interactions are mediated via M2 branes, which appear as strings on the M5 brane worldvolume. Thus, to consider tensor interactions calls for an appropriate accounting for the degrees of freedom carried by the self-dual string. From the point of view of amplitudes, this is difficult as the states are not particle-like. One approach is to approximate it as a tower of higher spin fields. However, as discussed in the previous section, allowing the self-dual tensor multiplet to couple to higher spins eventually lands one in a gravitational theory.    

\begin{figure}
\begin{center}
\includegraphics[scale=0.7]{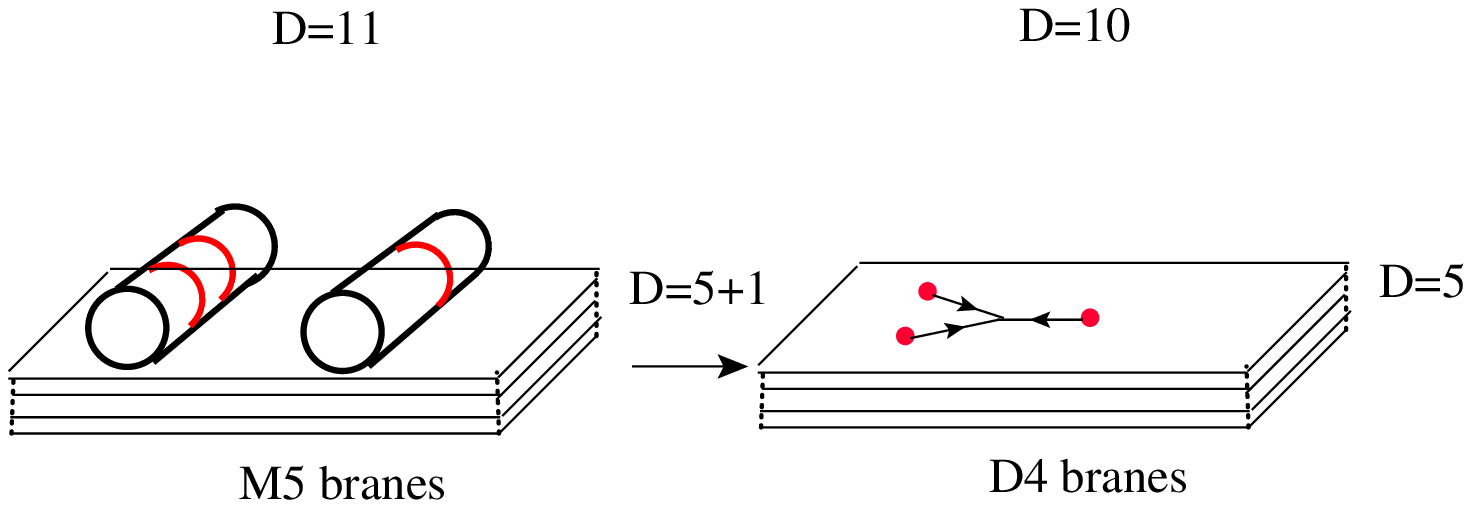}
\caption{ }
\label{talk}
\end{center}
\end{figure}

Another approach is to compactify the M5 branes on $\mathcal{M}^{4,1}\times {\rm S}^1$ and allow the self-dual strings to wrap around ${\rm S}^1$. In this configuration, the self-dual string modes will appear as massive particle states in five dimensions, as illustrated in fig.~\ref{talk}, and one can instead study five-dimensional massive amplitudes. Note that while the six-dimensional self-dual tensor field and the vector field have the same massless reduction to five dimensions, their massive reductions are distinct. In particular, they transform under different representations of the five-dimensional massive little group, which is isomorphic to the six-dimensional massless little group $SO(4)\sim SU(2)\times SU(2)$. Thus, in effect, we are looking for a massive extension of the five-dimenional super Yang-Mills amplitude that is ``chiral". This extension is then interpreted as describing the effective coupling of the KK modes that arises from the compactification of the self-dual string. The hope is that there are sufficient constraints to allow one to pin down a unique solution. 

We first give a brief review of five-dimensional massive spinor helicity.   

\subsection{Spinor helicity for D=$5+1$}
Both massive and massless five-dimensional spinor helicity can be easily understood as breaking the $SU^*(4)$ of the six-dimensional spinor helicity to its subgroup $USp(2,2)$, which is the covering group of the five-dimensional Lorentz group $SO(1,4)$. Using the $USp(2,2)$ metric, one can separate the six-dimenional massless vector into a five-dimensional vector plus a sixth component by separating the $USp(2,2)$ trace:
\eq
p^{(6)AB}=(p^{(6)AB}-\frac{1}{4}\Omega^{AB}p^{(6)CD}\Omega_{CD})+\frac{1}{4}\Omega^{AB}p^{(6)CD}\Omega_{CD}\equiv p^{AB(5)}+\Omega^{AB}m
\eqe
Here $\Omega_{AB}$ is the $USp(2,2)$ metric, $p^{AB(5)}$ is the traceless piece of the original $p^{(6)AB}$ and we have identified $p^{AB(6)}\Omega_{AB}=4m$. One can check that with these identifications the original six-dimensional massless condition becomes:
\eq
p^{(6)\mu}p^{(6)}_{\mu}=-\frac{1}{8}\epsilon_{ABCD}p^{(6)AB}p^{(6)CD}=p^{\mu(5)}p_{\mu}^{(5)}+m^2=0
\eqe
This equation follows from the fact that $\epsilon_{ABCD}$ can be rewritten as a combination of products of $USp(2,2)$ metrics as shown in eq.~(\ref{SpEp}), and uses
\eq
p^{\mu(5)}p_{\mu}^{(5)}=-\frac{1}{4}p^{(5)AB}p^{(5)}_{AB}\,.
\eqe
The momenta $p^{AB}$ are still expressed in bi-spinor form:
\eq
p^{AB}=\lambda^{Aa}\lambda^{B}\,_a
\eqe
The index $a$ is again the little group index, and the spinor is in the $(\mathbf{2},\mathbf{1})$ of the massive little-group $SU(2) \times SU(2)$. The anti-chiral spinor, which is a $(\mathbf{1},\mathbf{2})$, can also be defined in a $USp(2,2)$ covariant manner as:
\eq
\tilde{\lambda}_{i\dot{a}}\tilde{\lambda}_i\,^{\dot{a}}=\tilde{p}_{iAB} = -\frac{1}{2} \Omega_{A[B} \Omega_{CD]} \lambda_i^{Ca}\lambda^D_{ia} 
\label{SelfDualp}
\eqe 
Note that since there are no chiral spinors in odd dimensions, here chirality is really defined in terms of the little group representation. 

The $USp(2,2)$ metric allows us to contract the original six-dimensional (anti-)chiral spinors ($\tilde{\lambda}_{A\dot{a}}$) $\lambda_{a}^A$ to form five-dimensional Lorentz-invariant spinor inner products:
\eq
\langle i_a|j_b\rangle= \lambda^A_{ia}\lambda_{jAb} \qquad\qquad
[ \tilde{i}_{\dot{a}} | \tilde{j}_{\dot{b}} ]= \tilde\lambda^A_{i\dot{a}}\tilde\lambda_{jA\dot{b}}
\label{5DSpinIn}
\eqe
Note that since $p^{AB(6)}\Omega_{AB}=4m$, we have
\eq
\langle i^a|i_a\rangle =-4m\neq0.
\eqe
From eq.~(\ref{SelfDualp}), one can deduce 
\eq
\langle i^a|i_a\rangle=[\tilde{i}^{\dot{a}}|\tilde{i}_{\dot{a}}]\,.
\eqe
This simply reflects the fact that the mass is real. This analysis makes the massless limit transparent: massless spinor helicity corresponds to the additional constraint that the $USp(2,2)$ spinors form traceless vectors.

Unlike the six-dimensional spinor inner products, the new, five-dimensional spinor inner products are invertible for massive three-point kinematics. Indeed, from five-dimensional momentum conservation one has $p^{(5)}_i\cdot p^{(5)}_j=\frac{1}{2}(-m_k^2+m_i^2+m_j^2)$, while in terms of spinor inner products one has: 
\begin{equation}
p^{(5)}_i\cdot p^{(5)}_j=-\frac{1}{4}(2\det\langle i^a|j^b\rangle+\frac{1}{4} \langle i^a|i_a\rangle\langle j^b|j_b\rangle)
=  \,m_i \,m_j -\frac{1}{2}  \det\langle i^a|j^b\rangle\,\\
\end{equation}
Thus, the determinant is non-vanishing for generic masses. For the special case  $m_i+m_j+m_k=0$, the inverse is simply:
\begin{equation}
(\langle i_a|j_b\rangle)^{-1}=-\frac{\langle j^b|i^a\rangle}{4m_i m_j}
\label{invertm}
\end{equation}

Finally, we note that the momenta  $\tilde{p}_{iAB}$ defined above are inequivalent to the $USp(2,2)$ lowered $p_{AB} \equiv p^{CD} \Omega_{AC} \Omega_{BD}$:
\begin{equation}
\tilde{p}_{iAB} = -\frac{1}{2} \Omega_{A[B} \Omega_{CD]} \lambda_i^{Ca}\lambda^D_{ia} = 
\Omega_{AC}\Omega_{BD} p_i^{CD}-\frac{1}{2}\Omega_{AB}\langle i^{a}|i_a\rangle = p_{iAB} - 2m_i \Omega_{AB} \label{tildeandnot}
\end{equation}

We now consider the massless limit, i.e. $\langle i^a|i_a\rangle=0$. For three-point kinematics, we can choose a four-dimensional subspace to contain that spanned by the three momenta. One again obtains four-dimensional massless kinematics. With the choice $\Omega_{AB}={\rm diag}\{\epsilon^{\alpha\beta}, -\epsilon_{\dot{\alpha}\dot{\beta}} \}$ the new Lorentz invariants take a particularly simple form in our preferred four-space:
\begin{equation}
\label{5D4D}
\langle i_a | j_b \rangle 
= 
\left( \begin{array}{cccc} 
0                & 0               & \tilde{\lambda}_i^{(4)\dot{1}} & ^{(4)}\tilde{\lambda}_i^{(4)\dot{2}} \\
\lambda^{(4)}_{i1} & \lambda^{(4)}_{i2} & 0                                        & 0
\end{array} \right) 
\left( \begin{array}{rrrr} 
0 & -1 & 0 & 0 \\ 
1 & 0 & 0 & 0 \\
0 & 0 & 0 & -1 \\
0 & 0 & 1 & 0
\end{array} \right)
\left( \begin{array}{cc}
0 & \lambda^{(4)}_{j1} \\
0 & \lambda^{(4)}_{j2} \\
\tilde{\lambda}_{j}^{(4)\dot{1}} & 0 \\
\tilde{\lambda}_j^{(4)\dot{2}} & 0 
\end{array} \right) 
= 
\left( \begin{array}{cc}
[ij]   &0 \\
0 & -\langle ij\rangle
\end{array} \right)
\end{equation}
Similarly, with $\Omega_{AB}={\rm diag}\{\epsilon_{\alpha\beta}, -\epsilon^{\dot{\alpha}\dot{\beta}} \}$, we have 
\begin{equation}
\label{5D4D1}
[ \tilde{i}_{\dot{a}} | \tilde{j}_{\dot{b}} ]
= 
\left( \begin{array}{cc}
-[ij]   &0 \\
0 & \langle ij\rangle
\end{array} \right).
\end{equation}
Again, at this point one has to choose whether to set $\langle ij\rangle=0$ or $[ij]=0$.

Since both $\lambda$ and $\tilde{\lambda}$ can be consistently defined in a $USp(2,2)$ covariant way, one can continue to use the spinor inner product $\langle i_a|\tilde{j}_{\dot{a}}]$ in five dimensions. For three-point kinematics, when $\sum^3_{i=1}m_i=0$ holds, one has the full six-dimensional momentum conservation and $\langle i_a|\tilde{j}_{\dot{a}}]$ becomes rank one. For this case one again needs to rewrite $\langle i_a|\tilde{j}_{\dot{a}}]$ in terms of the $SU(2)$ variables $(u_a,\tilde{u}_{\dot{a}})$.

When the five-dimensional massive kinematics are obtained as a dimensional reduction of six-dimensional massless kinematics, then $\sum^3_{i=1}m_i=0$ is guaranteed. However, as we discuss in the next section, it is possible to have massive particle states in five dimensions that correspond to strings of the six-dimensional theory wrapping the compactified circle. In this case the five-dimensional mass is determined by the six-th component of momentum $p_{i}^5$ and the winding number of the string $n_i$ according to: 
\eq
|m_i|^2=|n_i+i\alpha p_{5i}|^2
\eqe
Here $\alpha$ is a dimensionful function of the radius of the circle $R$, and depends on the details of the quantization. For such configurations, one can again imbed the five-dimensional massive kinematics in six-dimensional massless ones by complexifying the six-dimensional momenta. In particular, we have:
\eq
p_j^{(6)}=(p_j^{(5)},n_j+i\alpha p_{5j}),\quad\quad \bar{p}_j^{(6)}=(p_j^{(5)},n_j-i\alpha p_{5j})
\eqe
The six-dimensional massless condition again gives the desired massive equation in five dimensions:
\eq
p_j^{(6)}\cdot \bar{p}_j^{(6)}=0\rightarrow p_j^{(5)}\cdot p_j^{(5)}+|m|^2=0
\eqe
Six-dimensional momentum conservation for both $p_j$ and $\bar{p}_j$ now implies five-dimensional momentum conservation as well as conservation of the winding mode $n_i$ and the sixth momentum component $p_{5j}$. Because the six-momentum is complex, one no longer has the reality condition in eq.~(\ref{SelfDualp}), i.e.  $\bar{p}_{AB}\neq\frac{1}{2}\epsilon_{ABCD}p^{CD}$. For three-point kinematics, while the five-dimensional masses no longer satisfy mass conservation, the six-dimensional massless condition coupled with momentum conservation implies that the spinor inner products $\langle i_a|\bar{j}_{\dot{a}}]$\footnote{Now the spinors $\lambda$ and $\bar{\lambda}$ are defined directly from $p_i^{AB}$ and $\bar{p}_{iAB}$, respectively, without reference to self-duality equations.} are still rank 1 and one can use the $SU(2)$ variables as before. In the five-dimensional notation, mass conservation is now replaced with:
\eq
\sum_i \langle i^a|i_a\rangle=0,\quad\quad \sum_i [ \bar{i}^{\dot{a}}|\bar{i}_{\dot{a}}]=0
\label{conservation}
\eqe
As the massive states discussed here are considered as arising from a compactification of a six-dimensional theory, we impose eq.~(\ref{conservation}) in addition to five-dimensional momentum conservation. This ensures that the sixth momentum component and winding number are conserved. 
\subsubsection{External line factors of the KK modes:}
The polarization tensors, or external-line factors, for the KK modes can simply be deduced from their six-dimensional counterpart given in eq.(\ref{pol}). A simplification arises due to the breaking of Lorentz symmetry down to five-dimensions: some of the components become pure gauge, and can be set to zero by a suitable choice of gauge, or equivalently a suitable choice of reference vector $\mu$. This can be seen by considering the abelian gauge transformation of the KK modes  for the self-dual tensor~\cite{Lambert:2010iw,Ho:2011ni}:
\eq
\delta B^{\rm kk}_{\hat{\mu}\hat{\nu}}=B^{\rm kk}_{\hat{\mu}\hat{\nu}}+\partial_{[\hat{\mu}}\Lambda^{\rm kk}_{\hat{\nu}]},\quad\delta B^{\rm kk}_{\hat{\mu}5}=B^{\rm kk}_{\hat{\mu}5}+\partial_{\hat{\mu}}\Lambda^{\rm kk}_{5}-\partial_{5}\Lambda^{\rm kk}_{\hat{\mu}}
\label{algebra}
\eqe
where $\hat{\mu}=0,1,2,3,4$. First of all, the gauge parameter $\Lambda^{\rm kk}_{5}$ is redundant as the above transformation rules are unmodified by $\delta\Lambda_{\hat{\mu}}^{\rm kk}=\partial_{\hat{\mu}}\lambda^{\rm kk},\,\delta\Lambda_{5}^{\rm kk}=\partial_{5}\lambda^{\rm kk}$. The remaining parameter $\Lambda^{\rm kk}_{\hat\mu}$ can be used to completely gauge away $B^{\rm kk}_{\hat{\mu}5}$ and hence the on-shell degrees of freedom are carried by $B^{\rm kk}_{\hat{\mu}\hat{\nu}}$. The self-duality relation for $B^{\rm kk}_{\hat{\mu}\hat{\nu}}$ now reads:
\eq
B^{\rm kk}_{\hat{\mu}\hat{\nu}}=\frac{1}{3!}\epsilon_{\hat{\mu}\hat{\nu}\hat{\rho}\hat{\sigma}\hat{\tau}}\frac{\partial^{\hat{\rho}}}{\partial_5}B^{\rm kk\hat{\sigma}\hat{\tau}}\,.
\label{selfdualKK}
\eqe
The relation in eq.(\ref{selfdualKK}) reduces the degrees of freedom carried by $B^{\rm kk}_{\hat{\mu}\hat{\nu}}$ down to three. To see this, note that one can choose the field to be in the rest frame, then the self-duality equation would indicate $B^{\rm kk}_{\hat{0}\hat{i}}=0$ and $B^{\rm kk}_{\hat{i}\hat{j}}=\frac{\epsilon_{\hat{i}\hat{j}\hat{k}\hat{l}}}{2}B^{{\rm kk} \hat{k}\hat{l}}$, where $\hat{i}=1,2,3,4$ and we have used $\partial^{\hat{0}}B^{\rm kk}=\partial^{5}B^{\rm kk}=m$.

We now consider the same story from the viewpoint of polarization tensors. The six-dimensional polarization tensors in eq.~(\ref{pol}) can be decomposed as
\eq
\tilde\epsilon^{AC}\,_{ab}(\mu,p_i)=\frac{\,^{A}|i_{(a}\rangle\langle i_{b)}|_B\mu^{BC}}{s_{i,\mu}}\rightarrow \left(\frac{\,^{[A}|i_{(a}\rangle\langle i_{b)}|_B\mu^{BC]}}{s_{i,\mu}},\;\;\frac{\,^{(A}|i_{(a}\rangle\langle i_{b)}|_B\mu^{BC)}}{s_{i,\mu}}\right),
\eqe
where we have used the $USp(2,2)$ metric to raise and lower the indices. The anti-symmetric and traceless piece corresponds to the polarization "vector" of $B_{5\hat{\mu}}$ while the symmetric piece corresponds to that of $B_{\hat{\mu}\hat{\nu}}$. One can immediately see that by choosing $\mu^{AB}\sim\Omega^{AB}$, the anti-symmetric piece vanishes while the symmetric piece simply becomes: 
\eq
\left.\tilde\epsilon^{AC}\,_{ab}(\mu,p_i)\right|_{\mu^{AB}=\Omega^{AB}}=\frac{\,^{A}|i_{(a}\rangle\langle i_{b)}|^{C}}{m_i}\,
\label{PolTens}
\eqe  
We see that indeed the anti-symmetric piece of the six-dimensional polarization tensor becomes pure gauge and after gauging it away, the remaining symmetric piece does not depend on any remnant reference spinor, indicating that there is no remnant gauge symmetry.
\subsubsection{ Three-point kinematics with one massless leg}
We now consider the special case of three-point kinematics when one of the legs is massless. In this case one has for the massless leg (say, leg 1):
\eq
\tilde{p}_{1AB}=p_{1AB}\quad \Rightarrow \quad \lambda_{1Aa}=i\tilde\lambda_{1A a}
\label{MassLessIdent}
\eqe
The two spinors are now identified because the massive little group $SU(2) \times SU(2)$ has been reduced to the diagonal $SU(2)$ of the massless vector. Note that $u_1$ and $\tilde{u}_1$ continue to be inequivalent, as they are defined with respect to different chirality spinors:
\eq
\langle 1_a|\tilde{2}_{\dot{a}}]=u_{1a}\tilde{u}_{2\dot{a}},\quad \langle 2_b|\tilde{1}_{a}]\equiv -u_{2b}\tilde{u}_{1a}=-i\langle 2_b|1_a\rangle\,
\eqe  
In particular, $\tilde{\lambda}_i\neq\lambda_i$ for $i=2,3$ since these are still massive legs and we can identify $\tilde{u}_{1a}\equiv i\tilde{u}_{1\dot{a}}$. This implies that we have the following new invariants:
\eq
(w_1^a \tilde{u}_{1a}, u_1^a \tilde{w}_{1a},w_1^a \tilde{w}_{1a}, u_1^a \tilde{u}_{1a})\,
\eqe

Some of the above invariants can be expressed in terms of spinor invariants of the remaining two legs. A useful example can be derived as follows:
\eqa
\nonumber u_{1}^{a}\langle 1_a|2_b\rangle&=&u_{2}^{a}\langle 2_a|2_b\rangle=2m_2u_{2b}\\
\nonumber&=&-iu_{1}^{a}[ 1_a|2_b\rangle=iu_{1}^{a}\tilde{u}_{1a}u_{2b}\\
\rightarrow&&u_{1}^{a}\tilde{u}_{1a}=-2im_2
\eqae
This tells us that we can express $\tilde{u}_1^{b}$ in the basis of $u_1,w_1$ as
\eq
\tilde{u}_1^{a}=\gamma u_1^a-i2m_2w_1^a,
\label{TildeU}
\eqe
where $\gamma$ is an unfixed parameter. It is non-zero since we expect that in the limit $m=0$, we should have $\tilde{u}_1^{b}=u_1^{b}$. Equivalently, one can define $\gamma$ as:
\eq
\gamma\equiv \tilde{u}_1^{a}w_{1a}
\eqe
Note the by choosing an explicit value for $\gamma$ we have fixed the $b$-shift ambiguity of $w_{1a}$, so the requirement of $b$-shift invariance in eq.~(\ref{bShift}) is reduced down to $(b_2,b_3)$, satisfying $b_2+b_3=0$. Since $\gamma$ scales as $\alpha^{-2}$, as long as the amplitude is invariant under $\alpha$-scaling, it will not depend on the explicit value of $\gamma$ if $\gamma\neq0$.
\subsection{Supersymmetry constraint with five-dimensional three-point kinematics}
By reducing the kinematics to five dimensions, the hope is that the reduced Lorentz symmmetry allows us to utilize as building blocks for the amplitude new Lorentz invariants, which were previously not allowed in the manifest six-dimensional analysis. We have already discussed the new invariants $\langle i_a|j_b\rangle$ and $[i_{\dot{a}}|j_{\dot{b}}]$, built from bosonic spinors and defined in eq.~(\ref{5DSpinIn}). For invariants that involve fermions, we note that 
\begin{eqnarray}
\nonumber 
\lambda_1^{Aa}\eta_{1a}\lambda_{2A}^b&=&2m_2\mathbf{w}_1u_2^b-\mathbf{u}_1w_{1a}\langle 1^a|2^b\rangle\\ \nonumber 
\lambda_1^{Aa}\eta_{1a} \lambda_{2A}^{a}\eta_{2a}
 &=&\left(\mathbf{u}_{1}\mathbf{u}_{2}w_{1c}\langle 1^c|2^d\rangle w_{2d}+2m_2\mathbf{w}_{1}\mathbf{u}_{2}-2m_1\mathbf{u}_{1}\mathbf{w}_{2}\right)\,,
\end{eqnarray}
so all Lorentz invariants involving inner products of supermomenta $\lambda_i^{Aa}\eta_{ia}$ can always be rewritten as polynomials of $\mathbf{u}$ and $\mathbf{w}$. Thus, in five dimensions, the three-point superamplitude is again a polynomial of the variables $(\mathbf{u}_i,\; \mathbf{w}_i)$.

Since the five-dimensional supersymmetry constraint $Q^A \mathcal{A}_3=0$ is identical to the six-dimensional version, one simply identifies the $SU(4)$ index as $USp(2,2)$ to conclude that the solutions to the supersymmetry constraints are again
\eqa
\nonumber\mathcal{N}=(1,0)&:&\quad \mathcal{A}_3\sim \Delta(Q)f \\
\nonumber\mathcal{N}=(2,0)&:&\quad \mathcal{A}_3\sim \Delta(Q)\Delta(\hat{Q})g\,,
\eqae
where $f$ and $g$ must be $b$-shift invariant and scale as $\alpha^{-1}$ and $\alpha^{-2}$, respectively. These are the same constraints as in six-dimensions. The only twist of the story now is that $f$ and $g$ can depend on more general, five-dimensional Lorentz invariants. 

We now look for three-tensor KK multiplet interactions. For $\mathcal{N}=(1,0)$ supersymmetry, this requires $f$ to carry three chiral $SU(2)$ indices. Since $f$ must have scaling $\alpha^{-1}$, $f$ can take the form:
\eq
f\sim \left(\tilde{u}_{i\dot{a}}[i^{\dot{a}}|j^{\dot{b}}]\tilde{u}_{j\dot{b}}\right)u_{ia}\langle i^a|j^b\rangle\langle l^c|k^d\rangle
\eqe
However, using eq.~(\ref{Ueq}), we see that this vanishes. Other possibilities require the inclusion of $w$ variables and can be ruled out from $b$-shift invariance. For $\mathcal{N}=(2,0)$, three tensor multiplet interaction requires $g$ to carry no little group index and scale as $\alpha^{-2}$. Due to similar arguments as above we can see that there are no solutions. 

We conclude that:  \textit{five-dimensional Lorentz invariance and little group covariance forbids three massive self-dual tensor mulitplet interactions with $\mathcal{N}=(1,0)$ or $\mathcal{N}=(2,0)$ supersymmetry}.\footnote{Although one can now write down three self-dual tensor interactions with no supersymmetry: $\frac{\langle1_{a)}|2_{(b}\rangle\langle2_{c)}|3_{(d}\rangle\langle 3_{e)}|1_{(f}\rangle}{m_1m_2m_3}$.}

We next turn to couplings involving two massive self-dual tensors. We find that in five dimensions, two massive tensors can now be coupled to a massive vector multiplet with $\mathcal{N}=(1,0)$ supersymmetry:
\eq
\framebox[10cm][c]{$\mathcal{N}=(1,0):\quad \mathcal{A}_3\sim \Delta(Q)\tilde{u}_{1\dot{a}}\langle 2_b|3_c\rangle\,$}
\label{N10}
\eqe 
We use $\sim$ to indicate that the amplitude is defined up to factors of $m_i$. Recall that such interaction was not allowed with manifest six-dimensional Lorentz invariance, which we attributed, in subsection~\ref{WhatsWrong}, to the incompatibility of the linear gauge transformations required for gauge invariance of the amplitude, and those of the free action. Our results indicate that the scattering of an $\mathcal{N}=(1,0)$ vector and two tensor multiplets is now possible in five dimensions. One way of understanding this result is that due to the breaking of Lorentz invariance down to five dimensions, it is now possible to reduce the gauge symmetry by gauging away some irreducible pieces of the fields, i.e. $B^{\rm kk}_{\hat{\mu}5}$. The remaining gauge symmetry, at the linear level, can now be made consistent with that of the free action. It would be interesting to construct an action that exemplifies this scenario.   

We now turn to three-point kinematics with one leg massless. This corresponds to considering the scattering of five-dimensional super-Yang-Mills with the KK modes, i.e. we are considering five-dimensional massless vector multiplet coupled to a massive self-dual tensor multiplet. For $\mathcal{N}=(1,0)$ supersymmetry, one can simply take leg 1 in eq.~(\ref{N10}) to be massless. For $\mathcal{N}=(2,0)$ one has the following possibility
\eq
\mathcal{N}=(2,0):\quad \mathcal{A}_3\sim \Delta(Q) \Delta(\hat{Q})(\tilde{u}_{1}^aw_{1a})\,,
\eqe 
where again $\sim$ indicates that the amplitude is defined up to factors of $m$, which we will fix in the next subsection. Note that in contrast to $\mathcal{N}=(1,0)$, we have to reduce the $b$-shift invariance of the amplitude in eq.~(\ref{bShift}) to only $(b_2,b_3)$ by fixing a non-zero ``gauge" for $ \gamma\equiv\tilde{u}_{1}^aw_{1a}$.
\subsection{$\mathcal{N}=(2,0)$ amplitudes}
In this subsection we consider $\mathcal{N}=(2,0)$ amplitudes that have a natural understanding as describing the interactions of the KK modes when multiple M5 branes are wrapped on a circle. From the previous discussion, we see that $\mathcal{N}=(2,0)$ amplitudes can be determined up to factors of $m_i$. Here we will fix this ambiguity by determining the mass dimension of the amplitude. An interesting constraint is the expected $S$-duality that should emerge if one further compactifies the theory on $\mathcal{M}_4\times T^2$.

\subsubsection{Constraint from S-duality}
In ref.~\cite{Douglas:2010iu} Douglas showed that by including the contribution of the two towers of massive KK modes in the four-dimensional $\mathcal{N}=4$ super-Yang-Mills one-loop four-gluon amplitude, one obtains an S-duality invariant term for the effective action if one  assumes that the effective coupling of the KK modes to a massless gluon is independent of their winding number~\cite{Douglas:2010iu}:
\eq
\Delta\mathcal{L}\sim C_8L^4tr F^4,\quad C_8=\sum_{n_5,n_6\neq0}\left(\frac{{\rm Im}\tau}{|n_6\tau+n_5|^2}\right)^2
\label{SDual}
\eqe
 Here $L^2$ stands for the volume of the $T^2$ and $n_5,n_6$ are the quantum numbers for the torus. More precisely, this function was obtained by taking the $\mathcal{N}=4$ one-loop four-point amplitude, which is simply $stA_4^{tree}$ times a scalar box integral, and replacing the propagators in the box integrals with massive ones to take into account of the effect of the KK-modes. Taking the large mass limit ($L^2 \ll 1$) and summing over the two towers of quantum numbers $n_5,n_6$ one obtains eq.~(\ref{SDual}). 

An implicit assumption in this computation is that the one-loop amplitude generated by the effective coupling between KK modes and the massless gluon is still a simple box integral with no triangle or bubble contributions. In four dimensions, this is perfectly valid since the KK states appear as massive vectors and it is known that maximal super-Yang-Mills with massive vectors respects a Dual Conformal symmetry~\cite{Alday:2009zm,CaronHuot:2010rj,Dennen:2010dh}, which dictates that the four-point one-loop amplitude is still expressed in terms of a scalar box integral.  

Now consider the same process in five dimensions, i.e. we consider the five-dimensional coupling between the zero modes and the massive KK-modes. The one-loop four-point amplitude of this process will encode the four-dimensional daughter amplitude discussed above, since if we set the four external legs in a four-dimensional subplane, then kinematically the process is the same except that the integration of the loop momenta in the fifth dimension for the internal states becomes a sum over one of the towers in four-dimensions. The fact that the five-dimensional one-loop four-point amplitude will encode the four-dimensional daughter implies that the former must be a scalar box integral as well, since the presence of any triangle or bubble integrals would imply that such integrals would have appeared in the four-dimensional analysis. Note that this is a non-trivial constraint in five dimensions since now the KK-modes are not massive vector multiplets but rather massive tensors, and the Dual Conformal symmetry for massive maximal super-Yang-Mills can no longer be utilized.

We now argue that this ``no-triangle" constraint can be translated into the constraint that the three-point amplitude must have mass-dimension $\leq1$.\footnote{A much simpler argument, although a bit hand waving, would be that the two-KK interaction with a massless gluon must reduce to the two-massive vector and one gluon interaction in four-dimensions. Hence it must be a mass-dimension 1 in five dimensions. } It can be shown, via Passarino-Veltman integral reductions~\cite{Passarino:1978jh}, that all one-loop four-point amplitudes can be cast into a basis of scalar integrals that include bubbles, triangles, and boxes along with purely rational functions. The integral reduction reduces any four-point integrals with numerators of degree $m$ in loop momentum $\ell$, denoted $N^m(\ell)$, into a basis of scalar box integrals and triangles with numerators of degree $m-1$: 
\eq
I_4\left[N^m(\ell)\right]\rightarrow \sum_i c_i I^i_4[N^0]+\sum_{j} I^j_3[N^{m-1}(\ell)]\,
\eqe
Iterating the same reduction on triangles and bubbles one obtains the previously stated scalar integral basis. If one assumes that the three-point interaction is an $n$-derivative coupling in the effective action, then the Feynman rules will give four-point one-loop integrands that have at most $4n$-powers of loop momenta in the numerator. Since we are considering a maximal supersymmetric theory, the one-loop amplitude must be proportional to the supermomenta delta function which has mass dimension $4$ and is a function of external momenta. This then implies that the four-point integrand can have numerators that have at most $4n-4$ powers of loop momenta. But the presence of \textit{any} loop momenta will produce scalar triangle integrals via integral reductions. Thus, the constraint that the one-loop four-point amplitude only contains box integrals can only hold if $n\leq1$. This means that the three-point interaction must be at most $1$-derivative, or equivalently, \textit{the three-point amplitude must have mass-dimension $\leq1$.}

\subsubsection{Two KK-tensor and one massless vector}
The above analysis fixes the $\mathcal{N}=(2,0)$ three-point amplitude involving two KK-tensors and one massless gluon. Since the interaction can only have mass dimension $\leq1$ and it should be independent of $n_6$ (and, as a consequence, of $m=n_6/R$), the amplitude is simply:
\eq
\framebox[10cm][c]{$\mathcal{N}=(2,0):\quad \mathcal{A}_3=\Delta(Q) \Delta(\hat{Q})(\tilde{u}_{1}^aw_{1a})\,$\label{BfieldBfield}}
\eqe 
Note that if we take the five-dimensional massless limit, which corresponds to identifying $u_i=\tilde{u}_i,\;w_i=\tilde{w}_i$, the above amplitude simply reduces to that of five-dimensional $\mathcal{N}=4$ sYM. This implies that five-dimensional sYM not only captures most of the KK modes of the (2,0) theory~(see \cite{Rozali:1997cb} and more recently \cite{Lambert:2010iw, Douglas:2010iu}), but some of their dynamics as well. We note that while the massless limit of compactified six-dimensional $\mathcal{N}=(1,1)$ sYM also gives maximal sYM in five dimensions, its massive extension is different. It is given by:
\eq
\mathcal{A}_3^{\mathcal{N}=(1,1)}=\delta^5(P)\Delta(Q)\Delta(\tilde{Q})
\eqe 

The fact that the self-dual tensor KK modes interact through a zero mode is reminiscent of the recent proposal for an effective bosonic action of multiple M5 branes in $D=5+1$~\cite{Ho:2011ni}, where the three-point interaction is mediated through vector zero modes.

\subsubsection{Pure KK interactions of two spin-3/2 and one tensor}
Since the KK-modes couple to the zero mode with a one-derivative coupling, it is natural to expect all other three-point interactions to have the same mass-dimension as there is only one coupling constant, set by the scale of the compactification radius. Interestingly, we have the following $\mathcal{N}=(2,0)$ amplitude that satisfies this criterion: 
\eq
\framebox[10cm][c]{$\mathcal{A}_3=\delta^5(P)\Delta(Q)\Delta(\hat{Q})\frac{\tilde{u}_{j\dot{a}}\tilde{u}_{k\dot{b}}}{m_i}g^{ijk}$\label{Spin3/2}}
\eqe
Here $i\neq j\neq k$, $g^{ijk}=g^{ikj}$ and $m_i$ is the mass of the self-dual tensor multiplet. Note that besides the factor $\frac{1}{m_i}$, the amplitude has the same form as eq.~(\ref{SelfDualSpin}), whose coupling in terms of component fields we derived in eq.~(\ref{BAA2}). Including the $\frac{1}{m_i}$ factor, one can derive the component field interaction in five dimensions. We begin with
\eq
\mathcal{L}_3\sim\epsilon_{\mu\nu\rho\sigma\tau\delta}\left(\frac{1}{\partial_5}B^{i\mu\nu}\right)F^{j\rho\sigma}F^{k\tau\delta}g^{ijk}=-\epsilon_{\mu\nu\rho\sigma\tau\delta}\left(\frac{\partial^\rho}{\partial_5}B^{i\mu\nu}\right)A^{j\sigma}F^{k\tau\delta}g^{ijk}\,.
\eqe
Going to five dimensions and taking into account that on-shell $B^{\rm kk}_{5\hat\mu}=0$, we arrive at\footnote{There is an additional term $\epsilon_{\hat\mu\hat\nu\hat\rho\hat\sigma\hat\tau}\left(\frac{\partial^{\hat\rho}}{\partial_5}B^{{\rm kk}i\hat\mu\hat\nu}\right)A^{j\hat\sigma}\partial^{5}A^{k\hat\tau}g^{ijk}$ which can be shown to vanish by using momentum conservation for $k_5$ and the fact that $g^{ijk}=g^{ikj}$.} 
\eqa
\mathcal{L}_3&\sim&-\frac{\epsilon_{\hat\mu\hat\nu\hat\rho\hat\sigma\hat\tau}}{6}\left[B^{{\rm kk}i\hat\mu\hat\nu}A^{j\hat\rho}F^{k\hat\sigma\hat\tau}-2\left(\frac{\partial^{\hat\rho}}{\partial_5}B^{{\rm kk}i\hat\mu\hat\nu}\right)A^{j5}F^{k\hat\sigma\hat\tau}\right]g^{ijk}\,.
\eqae
Using the self-duality equation for the KK modes in eq.~(\ref{selfdualKK}), one sees that the above three-point interaction can be written as:
\eqa
\mathcal{L}_3&\sim&\left[-\frac{\epsilon_{\hat\mu\hat\nu\hat\rho\hat\sigma\hat\tau}}{6}B^{{\rm kk}i\hat\mu\hat\nu}A^{j\hat\rho}F^{k\hat\sigma\hat\tau}+\frac{B^{{\rm kk}i}_{\hat\sigma\hat\tau}}{3}A^{j5}F^{k\hat\sigma\hat\tau}\right]g^{ijk}\,
\eqae
Note that although this is also a vector-vector-tensor interaction, it is different from the $B\wedge F\wedge F$ interaction for six-dimensional $\mathcal{N}=(2,0)$ supergravity. The superamplitude in eq.~(\ref{Spin3/2}) describes the supersymmetrization of the above interaction and one can identify the factor $\frac{1}{m_i}$ as coming from the external line factor for the massive tensor fields given in eq.~(\ref{PolTens}).

\subsubsection{ Self-dual string on  $\mathcal{M}_5\times {\rm S}_1$}
It will be enlightening to identify the above massive $\mathcal{N}=(2,0)$ multiplets in the context of 1/2- and 1/4-BPS representations of the five-dimensional massive SUSY algebra. We now give a brief review of the massive KK states that arise from the self-dual six-dimensional tensionless string. For a more detailed discussion see ref.~\cite{Lambert:2010iw}. 

We begin by considering the most general (2,0) supersymmetry algebra consistent with the 11-dimensional projection $\Gamma_{012345}Q_\alpha=-Q_\alpha$~\cite{Howe:1997et, D'Auria:2000ec,Lambert:2010iw}: 
\eq
\{Q_\alpha,Q_\beta\}=P_m(\Gamma^mC^{-1})^{-}_{\alpha\beta}+Z_m^I(\Gamma^m\Gamma^I C^{-1})^{-1}_{\alpha\beta}+Z_{mnp}^{IJ}(\Gamma^{mnp}\Gamma^{IJ}C^{-1})^-_{\alpha\beta}
\eqe
Here $m,n=0,..., 5$, $I,J=6,....,10$, $C=\Gamma_0$, $Z_{mnp}^{IJ}$ is self-dual and the superscript $^-$ indicates a projection to the negative eigenstates of $\Gamma_{012345}$. Self-dual strings wrapping a cycle will appear in five dimensions as massive particle-like states. These can be interpreted as BPS states of the (2,0) supersymmetry with mass $M$, momentum $P_5$ along $x_5$ (the M5 brane direction not included in the D4 brane worldvolume) and central charge $Z_5^I$ corresponding to the electric central charge of five-dimensional super-Yang-Mills. 

In the unbroken phase $Z_5^I=0$, the KK modes are packaged into the massive self-dual tensor multiplet, with the self-dual tensor an R-symmetry singlet.  This multiplet is uncharged. Let us consider the broken phase of the theory, where one of the scalars of the $\mathcal{N}=(2,0)$ develops a VEV. Following~\cite{Lambert:2010iw} we choose the scalar VEV to be $\langle\phi^6\rangle$. This gives a non-vanishing electric central charge $Z^6_5\neq0$ and the algebra becomes:
\eq
\{Q_\alpha,Q_\beta\}=(M+P_5\Gamma^{50}+Z_5^6\Gamma^5\Gamma^6\Gamma^0)^{-}_{\alpha\beta}
\eqe  
The BPS condition then corresponds to the zero eigenstates of $M+P_5\Gamma^{50}-Z_5^6\Gamma^{50}\Gamma^6$. Since $\Gamma^{50}$ and $\Gamma^{50}\Gamma^6$ commute, a BPS state is an eigenstate of both projection operators and hence preserves only a quarter of the supersymmetry. The broken 12 supercharges can be further separated according to their representation in the massive little group $SU(2) \times SU(2)$, with 8 supercharges forming chiral spinors $(\hat{\mathbf{2}},\hat{\mathbf{1}})$ and the remaining 4 forming anti-chiral spinors $(\hat{\mathbf{1}},\hat{\mathbf{2}})$. We have used hatted numbers to indicate the representation under the R-symmetry to differentiate from the little group. 

The fermionic field content of the resulting self-CPT multiplet comprises a spin-3/2 fermion $S_{(ab)}\,^{\dot{a}}$, 5 anti-chiral spinors $\psi^{\dot{a}}$, and 8 chiral spinors $\chi^a$.\footnote{Here, for consistency, we choose the opposite chirality convention to ref.~\cite{Lambert:2010iw}.} Their representation under the little group and R-symmetry group is given by:
\eqa
\nonumber S_{(ab)}\,^{\dot{a}}&:& \quad(\mathbf{3},\mathbf{2},\hat{\mathbf{1}},\hat{\mathbf{1}})\;\;[1]\\
\psi^{\dot{a}}&:&\quad(\mathbf{1},\mathbf{2},\hat{\mathbf{1}},\hat{\mathbf{1}})\oplus(\mathbf{1},\mathbf{2},\hat{\mathbf{2}},\hat{\mathbf{2}})\;\;[5]\\
\nonumber \chi^a &:&\quad(\mathbf{2},\mathbf{1},\hat{\mathbf{1}},\hat{\mathbf{1}})\oplus(\mathbf{2},\mathbf{1},\hat{\mathbf{2}},\hat{\mathbf{2}})\oplus(\mathbf{2},\mathbf{1},\hat{\mathbf{3}},\hat{\mathbf{1}})\;\;[8]
\eqae
The bosonic fields include two self-dual tensors $B_{\mu\nu}$, 4 vectors $F_a\,^{\dot{a}}$, and 10 scalars:
\eqa
\nonumber B_{(ab)}&:&\quad(\mathbf{3},\mathbf{1},\hat{\mathbf{2}},\hat{\mathbf{1}})\;\;[2]\\
F_a\,^{\dot{a}}&:&\quad(\mathbf{2},\mathbf{2},\hat{\mathbf{1}},\hat{\mathbf{2}})\oplus(\mathbf{2},\mathbf{2},\hat{\mathbf{2}},\hat{\mathbf{1}})\;\;[4]\\
\nonumber \phi&:&\quad(\mathbf{1},\mathbf{1},\hat{\mathbf{3}},\hat{\mathbf{2}})\oplus(\mathbf{1},\mathbf{1},\hat{\mathbf{2}},\hat{\mathbf{1}})\oplus(\mathbf{1},\mathbf{1},\hat{\mathbf{1}},\hat{\mathbf{2}})\;\;[10]
\eqae
All the above fields are complex. One can again use the on-shell supersymmetries of $\mathcal{N}=(2,0)$ to pack these KK modes into on-shell complex superfields:
\eqa
\nonumber  \Psi^{\dot{a}(+,+)} & = & \psi^{\dot{a}(+,+)}+\eta^{a}F_{a}\,^{\dot{a}(0,+)}+\hat{\eta}^{a}F_{a}\,^{\dot{a}(+,0)}+\eta^2\psi^{\dot{a}(-,+)}+(\eta^a\hat{\eta}_a)\psi^{\dot{a}(0,0)}+\hat{\eta}^2\psi^{\dot{a}(+,-)}\\
&+&\eta^{(a}\hat{\eta}^{b)}S_{(ab)}\,^{\dot{a}(0,0)}+\eta^2\hat{\eta}^aF_a\,^{\dot{a}(-,0)}+\hat{\eta}^2\eta^aF_a\,^{\dot{a}(0,-)}+\eta^{2}\hat{\eta}^{2}\psi^{\dot{a}(-,-)}\\
\nonumber && \\
\nonumber\Phi^{(++,+)}\!&=&\!\phi^{(++,+)}+\eta^{a}\chi_{a}^{(+,+)}+\hat{\eta}^{a}\chi'^{(++,0)}_{a}+\eta^{2}\phi^{(+,+)}+\hat\eta^{2}\phi^{(++,-)}+(\eta^{a}\hat\eta_{a})\phi^{(+,0)}\\
&+&\eta_{(a}\hat\eta_{b)}B^{(ab)(+,0)}+\hat\eta^2\eta^{b}\chi_{b}^{(+,-)}+\eta^{2}\hat{\eta}^{b}\chi^{(0,0)}_{b}+\eta^2\hat\eta^{2}\phi^{(0,-)}\\
\nonumber && \\
\nonumber\Phi^{(0,+)}\!&=&\!\phi^{(0,+)}+\eta^{a}\chi_{a}^{(-,+)}+\hat{\eta}^{a}\chi^{'(0,0)}_{a}+\eta^{2}\phi^{'(--,+)}+\hat\eta^{2}\phi^{''0,-}+(\eta^{a}\hat\eta_{a})\phi^{'''(-,0)}\\
&+&\eta_{(a}\hat\eta_{b)}B^{(ab)(-,0)}+\hat\eta^2\eta^{b}\chi_{b}^{(-,-)}+\eta^{2}\hat{\eta}^{b}\chi^{'(--,0)}_{b}+\eta^2\hat\eta^{2}\phi^{''''(--,-)}
\eqae
We have labeled the fields with their charge under the $U(1)$ of each R-symmetry $SU(2)$ with units $1/2$; recall that $\eta$ and $\hat{\eta}$ carry charges $(+,0)$ and $(0,+)$, respectively. Note that the charges of $\Phi^{(++,+)},\Phi^{(0,+)}$ reflect the fact that the self-dual tensors are R-symmetry doublets. In contrast, for R-symmetry singlet self-dual tensors, such as in the preceding six-dimensional discussion, the superfield carries charges $\Phi^{(+,+)}$.

\subsubsection{Summary}
From the above analysis we find that by reducing the Lorentz group to five dimensions, new $\mathcal{N}=(2,0)$ amplitudes can be written down for interactions that have the mass dimension expected from a four-dimensional analysis. Since $\Delta(Q)\Delta(\hat{Q})$ carries $U(1) \times U(1)$ charge $(+++,+++)$, from little group indices one can deduce that the above two amplitudes correspond to:
\eqa
\nonumber 
\textrm{eq.~(\ref{BfieldBfield})}:\quad&\rightarrow& \langle\Phi^{(+,+;-m)}\Phi^{(+,+;m)}\Phi^{(+,+;0)}\rangle,\langle\Phi^{(++,+;-m)}\Phi^{(0,+;+m)}\Phi^{(+,+;0)}\rangle\\
\textrm{eq.~(\ref{Spin3/2})}:\quad&\rightarrow& \langle\Phi^{(+,+;m_1)}(i)\Psi^{(+,+;m_2)}_{\dot{b}}(j)\Psi^{\dot{c}(+,+;m_3)}(k)\rangle
\eqae  
Here we have included a third entry in the superscript of the superfield to indicate the quantum number of the six-th momentum components, which satisfy conservation rules in eq.~(\ref{conservation}); the $m$ can be complex.

\begin{figure}
\begin{center}
\includegraphics[scale=0.8]{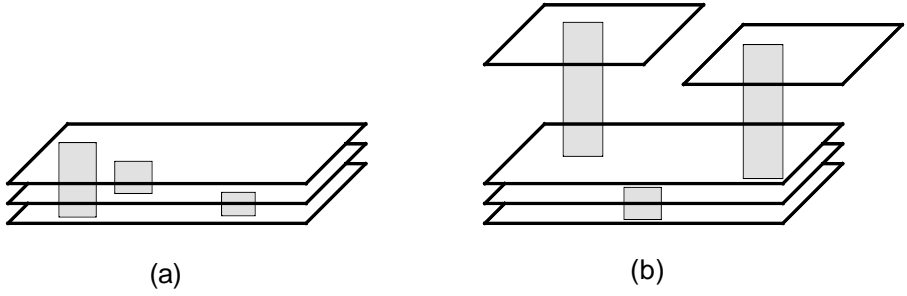}
\caption{Interactions in eq.~(\ref{BfieldBfield}) and eq.~(\ref{Spin3/2}) describe uncharged KK mode interaction as well as two-charged KK mode, one-uncharged interaction. They correspond to coincident and a few separated M5 branes, respectively.}
\label{M5branes}
\end{center}
\end{figure}

Thus, using simply five-dimensional super-Poincar\'{e} invariance for massive amplitudes, the requirement of S-duality of the four-dimensional descendant dictates that the participating multiplets are precisely the KK modes of the self-dual string. The case (\ref{BfieldBfield}) corresponds to either a three-self-dual tensor KK multiplet interaction, or a one-tensor, two-charged KK mode interaction. Since charged KK modes correspond to the theory in the broken phase, these two interactions apply to M5 brane configurations in diagrams (a) and (b) of fig.~\ref{M5branes}, respectively. For eq.(\ref{Spin3/2}), one has again a one-tensor, two-charged KK mode interaction, which corresponds to configuration (b) in fig.~\ref{M5branes}.

\subsection{The fermionic gauge algbera}
Before moving to the four-point amplitude, we first comment on the spin-$\frac{3}{2}$ state $S_{(ab)}\,^{\dot{a}}$, which has the same on-shell degrees of freedom as a gravitino, i.e. six. In ref.~\cite{Lambert:2010iw}, it was identified as a fermion with two spacetime vector indices, which satisfies the four-dimensional self-duality relation 
\eq
S^{ij}=\frac{1}{2}\epsilon^{\,ijkl}S_{kl},
\eqe 
where $i,j=0,1,2,3$. However, the fact that its massive little group representation has opposite chirality to the six-dimensional gravitino leads us to instead identify $S_{(ab)}\,^{\dot{a}}$ as a spin-3/2 particle. Furthermore, since $S_{(ab)}\,^{\dot{a}}$ is a singlet under the chiral R-symmetry, one can identify this spin-$3/2$ field as the gauge field for a local SUSY parameter 
\eq
\delta(\psi_\mu)^{\alpha A}=\partial_\mu \epsilon^{\alpha A}\,,
\label{newSUSY}
\eqe 
where the index $\alpha$ is the anti-chiral $Sp(2)$ R-symmetry index. The opposite chirality of the local fermionic symmetry allows one to close the gauge algebra to the vector gauge symmetry of the tensor instead of the graviton. We demonstrate this using the six-dimensional $SU^{*}(4)$ notation for simplicity; rewriting the result in five-dimensional $USp(2,2)$ notation is straightforward. 

In the $SU^*(4)$ notation, the two-form tensor is written as $B_A\,^B$, satisfying a traceless condition, $B_A\,^A=0$, hence giving a total of 15 components. The abelian self-dual and anti-self-dual field strength is given by 
\eq
{\rm (Self\;dual)}\;\;H_{(AB)}=\partial_{C(A}B_{B)}\,^C,\quad {\rm (Anti\;self\;dual)}\;\;H^{(AB)}=\partial^{C(A}B_{C}\,^{B)}\,.
\eqe
We propose the following local SUSY transformation: 
\eqa
\nonumber \delta B_A\,^B&=&\kappa\left[\epsilon^{C\alpha}(\psi_{\alpha AC})^B-\epsilon^{B\alpha}(\psi_{\alpha AC})^C-\frac{\delta^B_A}{2}\epsilon^{C\alpha}(\psi_{\alpha DC})^D\right]\\
\delta (\psi_{ AB})^{\alpha C}&=&\frac{1}{\kappa}\partial_{AB}\epsilon^{\alpha C}\,.
\eqae
Note that the indices on the spin-3/2 particle are all flat indices and $\kappa$ is a dimensionful parameter. Taking two commutations on the tensor field, one finds:
\eqa
[\delta_{\epsilon_2},\delta_{\epsilon_1}]B_A\,^B=\partial_{AC}[(\epsilon_{1})^{\alpha [C}(\epsilon_{2})_{\alpha}^{B]}]-\frac{\delta^B_A}{2}\partial_{DC}[(\epsilon_{1})^{\alpha C}(\epsilon_{2})_{\alpha}^{D}]\,
\eqae
Thus, the commutation of two fermionic local gauge transformations gives back the abelian vector gauge transformations of the two-form tensor, with the vector gauge parameter $\Lambda^{CB}=(\epsilon_{1})^{\alpha [C}(\epsilon_{2})_{\alpha}^{B]}$.\footnote{A tensor gauge transformation with parameter $\Lambda^{\mu}$ in $SU(4)$ notation takes the form
$$\delta B_A\,^C=\partial_{AB}\Lambda^{BC}-\frac{\delta^C_A}{4}\partial_{DB}\Lambda^{BD}.$$}

One can also verify that the free equations of motions are invariant under this local SUSY transformation. The free equations of motions are given by 
\eqa
\nonumber \partial^{AB}(\psi_{BC})^D=0, \quad H^{(AB)}=0\,.
\eqae
One can easily check that the vanishing of the anti-self-dual field strength is invariant under the local SUSY transformation due to the fermionic equation of motion.

In six dimensions, the dimensionful parameter $\kappa$ can only be related to the Plank length. However, since these are really five-dimensional massive particles, one now has a new scale set by the compactification radius, $R$. This allows us to have a local fermionic gauge symmetry in a non-gravitational theory.  Note that the presence of this fermionic gauge symmetry implies the existence of an additional global supersymmetry. Indeed, it is straightforward to see that the massive BPS states can be organized into an $\mathcal{N}=(2,1)$ multiplet, and hence supersymmetry is accidentally enhanced for the massive theory.   
\subsection{Constructing the four-point amplitude}
Equipped with the three-point amplitude, we can try to construct the four-point amplitude with six-dimensional BCFW recursion, now interpreted as five-dimensional massive BCFW. While we do not have an action to confirm the validity of the recursion, we can make consistency checks by requiring that one arrives at the same result irrespective of which pair of legs is chosen for the BCFW shift. Our aim here is to construct the two-KK, two-zero mode amplitude. 

\begin{figure}
\begin{center}
\includegraphics[scale=0.9]{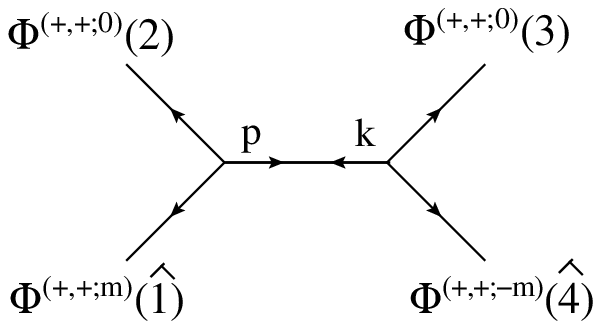}
\caption{ }
\label{New4pt}
\end{center}
\end{figure}
A detailed discussion of BCFW recursion using six-dimensional spinor helicity is given in refs.~\cite{cheung0, Dennen:2009vk}. As shown in fig.~\ref{New4pt}, we BCFW-shift legs 1 and 4 to obtain 
\eq
\mathcal{A}(\hat{1},2,3,\hat{4})=\frac{i}{s_{12}}\int d^2\eta d^2\hat{\eta}\mathcal{A}_3(\hat{1},2,p)\mathcal{A}_3(\hat{4},3,k),
\label{BCFW1}
\eqe
where legs $2$ and $3$ are the massless legs and
\eqa
\nonumber\mathcal{A}_3(\hat{1},2,p)&=&\Delta(Q_L)\Delta(\hat{Q}_L)\left( w_2^a\tilde{u}_{2a}\right)\\
\mathcal{A}_3(\hat{4},3,k)&=&\Delta(Q_R)\Delta(\hat{Q}_R)\left( w_3^a\tilde{u}_{3a}\right).
\eqae
The hatted bosonic variables in the above equations indicate shifted variables and  $s_{12}=2p_1\cdot p_2=(p_1+p_2)^2+m^2$. Recall that there is an $\alpha$-scale invariance of the three-point tree amplitudes on both sides of the BCFW bridge, which we denote by $\alpha_L$ and $\alpha_R$. Great simplification can be achieved if we gauge fix one of them, say $\alpha_L$, such that $\left( w_2^a\tilde{u}_{2a}\right)\times\left( w_3^a\tilde{u}_{3a}\right)$ is $1$. Thus, we are really computing 
\eq
\mathcal{A}(\hat{1},2,3,\hat{4})=\frac{i}{s_{12}}\int d^2\eta_P d^2\hat{\eta}_P\Delta(Q_R)\Delta(\hat{Q}_R)\Delta(Q_R)\Delta(\hat{Q}_R).
\eqe 
This computation is very similar to the computation of the four-point amplitude for $\mathcal{N}=(1,1)$ super-Yang-Mills theory in~\cite{Dennen:2009vk}. While in that case one has spinors and supercharges of both chiralities, in the explicit computation the two chiralities are done separately, giving the same result with the exchange of chiral and anti-chiral spinors. Details of the computation are relegated to Appendix \ref{last}; here we only display the final result:
\eq
\mathcal{N}=(2,0):\quad\mathcal{A}(\hat{1},2,3,\hat{4})=\frac{i\delta^4(Q_{full})\delta^{4}(\hat{Q}_{full})}{s_{12}s^{(6)}_{14}}
\label{4ptans}
\eqe
\eq
\textrm{with }\qquad \qquad 
\delta^4(Q_{full})=\frac{\epsilon_{ABCD}}{4!}Q^A_{full}Q^B_{full}Q^C_{full}Q^D_{full},\quad Q^A_{full}=\sum_{i=1}^4\;Q^A_{i}
\eqe
and a similar definition for $\delta^4(\hat{Q}_{full})$. Here $s^{(6)}_{14}=2p^{(6)}_1\cdot p^{(6)}_4=(p^{(5)}_1+p^{(5)}_4)$. This is essentially the same amplitude as the $\mathcal{N}=(1,1)$ super-Yang-Mills amplitude except now all the supercharges are chiral. This immediately leads to the result that the massless limit of this amplitude is exactly the massless $\mathcal{N}=4$ super-Yang-Mills amplitude in five dimensions. Note that the above result does not possess full six-dimensional Lorentz invariance; while it appears to be symmetric, our derivation relied on conservation of six-momentum, which implies that masses of adjacent pairs add to zero -- a condition that is not Lorentz invariant. 

Interestingly, it is known that the $\mathcal{N}=(1,1)$ four-point amplitude has a dual conformal symmetry~\cite{Dennen:2010dh}. To exhibit it, write $p_i=x_i-x_{i+1}$ and define for $n>3$:
\eq
\mathcal{A}_n=i\delta^5(P)\delta^4(Q_{full})\delta^{4}(\hat{Q}_{full})f_n\quad\, 
\eqe
Then under conformal inversion the functions $f_n$ transform covariantly \cite{Dennen:2010dh}:
\eq
I[f_n]=\prod_{i=1}^nx_i^2f_n
\eqe 
As we have shown, the $\mathcal{N}=(2,0)$ amplitude has the same $f_4$ as sYM, and hence also transforms covariantly under dual conformal symmetry. It would be interesting to see if this extends to higher point massive KK mode interactions.

\subsection{Toward $\mathcal{N}=(4,0)$ interactions}
The success in obtaining pure self-dual tensor amplitudes in five dimensions prompts us to look for other chiral amplitudes that have so far evaded us in six dimensions. One interesting object is amplitudes for the $\mathcal{N}=(4,0)$ multiplet.

This multiplet was first proposed in~\cite{Hull:2000zn} in an attempt to define a conformal invariant gravity theory in six dimensions that could serve as the strong coupling limit of the five-dimensional $\mathcal{N}=8$ supergravity. The interest in this theory has recently been revived~\cite{Chiodaroli:2011pp} as it may shed light on the UV divergent behavior of $\mathcal{N}=8$ supergravity following arguments similar to the $\mathcal{N}=(2,0)$ theory~\cite{Douglas:2010iu}. The  $\mathcal{N}=(4,0)$ contains 42 scalars, 27 self-dual tensors and one self-dual rank 4 tensor. Ref.~\cite{Hull:2000zn} showed that upon dimensional reduction to five dimensions, the states map perfectly to the maximal supergravity states. In particular, the rank 4 tensor, which transforms as a $(\mathbf{5},\mathbf{1})$ under the six-dimensional little group, becomes the five-component graviton. Thus, the $\mathcal{N}=(2,2)$ supergravity and the $\mathcal{N}=(4,0)$ theory arise from uplifting five-dimensional supergravity with respect to different duality frames.

Because we do not have a picture of what the $\mathcal{N}=(4,0)$ degrees of freedom should correspond to in terms of M-theory brane constructions, we have no intuition for the mechanism of the interactions. However, the theory has well defined asymptotic states since we have free equations of motion. Thus, one can again ask about the possible three-point amplitudes that involve these on-shell degrees of freedom. Requiring only super-Poincar\'{e} invariance, one immediately finds that there is one possible solution:
\eq
\mathcal{A}_{3}^{N=(4,0)}=\delta^5(P)\left[\prod_{i=1}^4\Delta(Q^i)\right]\left( w_k^a\tilde{u}_{ka}\right)^2\,
\eqe
It is easy to see that this is simply a product of $\mathcal{N}=(2,0)$ amplitudes. Hence, it has the desired property that the amplitude reduces to maximal supergravity in the massless limit. 

Note that the interaction is again only possible when one of the legs is massless. The KLT-like~\cite{Kawai:1985xq} structure of the above three-point amplitude hints at the possibility that the interacting mechanism of the $\mathcal{N}=(4,0)$ theory is related to the $\mathcal{N}=(2,0)$ theory via some open-closed string duality.

\section*{Acknowledgements}
Y-t would like to thank Robert Wimmer, Abhijit Gadde and Michael Douglas for very enlightening discussions. Y-t would also like to thank D. O'Connell, S. Ferrara, N. Lambert, C. Papageorgakis, M. Schmidt-Sommerfeld and Z. Bern for private communications and Mark Wise for the invitation as visiting scholar at Caltech. Part of this work was completed at the ``2011 Simons Summer Workshop in Mathematical Physics" at SCGP Stony Brook. The work of Y-t is supported by the US Department of Energy under contract DE-FG03-91ER40662. The work of BC and MR is supported by NSERC discovery grants.

\appendix
\section{Conventions and useful formulas\label{conventions}}
The contractions of $SU(2)$ indices follow
\eq
\psi^a=\epsilon^{ab}\psi_b,\;\;\psi_a=\psi^b\epsilon_{ab},\;\;\epsilon^{ab}\epsilon_{bc}=\delta^a_c
\label{metric}
\eqe
with 
\eq
\epsilon^{ab}=\left(\begin{array}{cc}0 & -1 \\1 & 0\end{array}\right),\quad\epsilon_{ab}=\left(\begin{array}{cc}0 & 1 \\-1 & 0\end{array}\right).
\eqe
We also have:
\eq
A_{[ab]}=A_{ab}-A_{ba}=\epsilon_{ab}A^c\,_c,\;\;A^{[ab]}=A^{ab}-A^{ba}=-\epsilon^{ab}A^c\,_c
\eqe
\subsection{Solutions for $u,w,\tilde{u},\tilde{w}$ where $\langle ij\rangle=0$\label{SquareSol}}
\vspace*{-12mm}
\begin{align}
u_{1a} = ( \,[2 3]^{-1},  \quad 0 \, ) \quad & \quad \tilde{u}_{1\dot{b}} = (\,[31][12], \quad 0 \, ) \nonumber \\
u_{2a} = ( \,[3 1]^{-1}, \quad 0\, ) \quad & \quad \tilde{u}_{2\dot{b}} = (\,[23][12],  \quad 0 \, ) \label{expl62} \\
u_{3a} = ( \,[1 2]^{-1}, \quad 0 \, ) \quad & \quad \tilde{u}_{3\dot{b}} = (\,[23][31],  \quad0 \, ) \nonumber
\end{align}
The pseudoinverses $w$ and $\tilde{w}$ take the form:
\begin{align}
w_{1a} = ( \,[2 3],  \quad 0 \, )  \quad & \quad 
\tilde{w}_{1\dot{b}} = (\, [1 2]^{-1}[3 1]^{-1},\quad 0 \, ) 
\nonumber \\
w_{2a} = ( \,[3 1], \quad 0\, ) \quad & \quad
\tilde{w}_{2\dot{b}} = (\, [1 2]^{-1}[2 3]^{-1},\quad 0 \, ) 
\label{explw2} \\
w_{3a} = ( \,[1 2], \quad 0 \, ) \quad & \quad
\tilde{w}_{3\dot{b}} = (\,[2 3]^{-1}[3 1]^{-1},\quad 0  \, )
\nonumber
\end{align}
\subsection{Breaking SU(4) to USp(2,2) }

The $USp(2,2)$ metric can be used to raise and lower indices in the same way as $SU(2)$:
\eq
\psi^A=\Omega^{AB}\psi_B,\;\;\psi_A=\psi^B\Omega_{BA},\;\;\Omega^{AB}\Omega_{BC}=-\delta^A_C
\eqe
The $SU(4)$ invariant tensor can be rewritten in terms of  $\Omega_{AB}$ as: 
\eq
\epsilon_{ABCD}=-\left(\Omega_{AB}\Omega_{CD}+\Omega_{AC}\Omega_{DB}+\Omega_{AD}\Omega_{BC}\right)
\label{SpEp}
\eqe

\section{BCFW recursion to eq.~(\ref{4ptans}) \label{last}}
The derivation of eq.~(\ref{4ptans}) follows similar computation as that done for four-point YM and sYM. We begin with the integration 
\eq
\int d^2\eta_P d^2\hat{\eta}_P\Delta(Q_R)\Delta(\hat{Q}_R)\Delta(Q_L)\Delta(\hat{Q}_L),
\label{Integrate}
\eqe
where 
\eqa
\nonumber\Delta(Q_R)&=&(\mathbf{u}_3\mathbf{u}_{\hat{4}}+\mathbf{u}_{\hat{4}}\mathbf{u}_{K}+\mathbf{u}_{K}\mathbf{u}_3)(\mathbf{w}_3+\mathbf{w}_{\hat{4}}+\mathbf{w}_{K})\\
\Delta(Q_L)&=&(\mathbf{u}_P\mathbf{u}_{\hat{1}}+\mathbf{u}_{\hat{1}}\mathbf{u}_{2}+\mathbf{u}_{2}\mathbf{u}_P)(\mathbf{w}_P+\mathbf{w}_{\hat{1}}+\mathbf{w}_{2})
\eqae
and $\mathbf{u}_{K}=u_K^a\eta_{Pa}$, $\mathbf{u}_P=u_P^a\eta_{Pa}$, and likewise for $\mathbf{w}_P, \mathbf{w}_K$. Because explicit integration gives a result that is proportional to $(u_P\cdot u_K, w_P\cdot w_K, u_P\cdot w_K, w_P\cdot u_K)$, we first evaluate these quantities explicitly.\footnote{The following derivation is based on private communication with Donal O'Connell. }

We first use the $b$-shift invariance of $w$ to fix $u_P\cdot w_K=u_K\cdot w_P=0$, and the integration result will be proportional to terms that have the form
\eqa
\nonumber &&\left[(u_P\cdot u_K)^2,\;(u_P\cdot u_K)(w_P\cdot w_K),\;(w_P\cdot w_K)^2,\;(u_P\cdot u_K),\;(w_P\cdot w_K),\;1\right]\\
\eqae 
Using the eq.~(\ref{Invert}) and eq.~(\ref{metric}), one can deduce 
\eq
(u_{P[a}w_{Pb]})(u_{K}\,^{[a}w_{K}\,^{b]})=\epsilon_{ab}\epsilon^{ab}=1\rightarrow (w_{P}\cdot w_{K})=\frac{1}{(u_{P}\cdot u_{K})}\,.
\eqe
Next, we consider the following object:
\begin{equation}
[ \hat{1}_{\dot{a}}|p_2p_{\hat{4}} |\hat{1}_{a}\rangle= -\tilde{u}_{\hat{1}\dot{a}}u^{d}_{2}\langle 2_{d}|p_{\hat{4}}|\hat{1}_{a}\rangle
=  \tilde{u}_{\hat{1}\dot{a}}u^{d}_{\hat{1}}\langle \hat{1}_{d}|p_{\hat{4}}|\hat{1}_{a}\rangle=-\tilde{u}_{\hat{1}\dot{a}}u_{\hat{1}a}s^{(6)}_{14}
\label{claytondown}
\end{equation}
In the last equality we used . On the other hand, we can also deduce:
\eqa
\nonumber [ \hat{1}_{\dot{a}}|p_2p_{\hat{4}} |\hat{1}_{a}\rangle&=& \tilde{u}_{1\dot{a}}u^{d}_{P}\langle P_{d}|p_{\hat{4}}|\hat{1}_{a}\rangle= i \tilde{u}_{1\dot{a}}u^{d}_{P}\langle K_{d}|p_{\hat{4}}|\hat{1}_{a}\rangle\\
\nonumber &=& i\tilde{u}_{1\dot{a}}(u_{P}\cdot u_{K})\tilde{u}_{\hat{4}\dot{b}}[\hat{4}^{\dot{b}}|\hat{1}_{a}\rangle= i\tilde{u}_{1\dot{a}}(u_{P}\cdot u_{K})\tilde{u}_{K\dot{b}}[K^{\dot{b}}|\hat{1}_{a}\rangle\\
&=&  u_{\hat{1}a}\tilde{u}_{\hat{1}\dot{a}}(u_{P}\cdot u_{K})(\tilde{u}_{P}\cdot\tilde{u}_{K})
\label{claytonup}
\eqae
Combining eqs.~(\ref{claytondown}) and (\ref{claytonup}) we arrive at:
$$(\tilde{u}_{P}\cdot\tilde{u}_{K})(u_{P}\cdot u_{K})=s^{(6)}_{14}\,.$$
Since we have only gauge fixed $\alpha_L$, we can use $\alpha_R$ to fix 
$$(\tilde{u}_{P}\cdot\tilde{u}_{K})=(u_{P}\cdot u_{K})=\sqrt{s^{(6)}_{14}}\,.$$

Using the above results for $(u_P\cdot u_K, w_P\cdot w_K, u_P\cdot w_K, w_P\cdot u_K)$, one can derive: 
\begin{align}
\nonumber \int  d^2 & \eta_P \Delta(Q_R) \Delta(Q_L) = 
\frac{1}{2\sqrt{s^{(6)}_{14}}} \mathbf{u}_{\hat{1}}\mathbf{u}_{2}\mathbf{u}_{3}\mathbf{u}_{\hat{4}} \,+ \\
\nonumber \frac{1}{4} & \Big( \mathbf{u}_{2}\mathbf{u}_{\hat{4}}\eta_3^2
-\mathbf{u}_{\hat{1}}\mathbf{u}_{\hat{4}}\eta_3^2
+\mathbf{u}_{\hat{1}}\mathbf{u}_{3}\eta_{\hat{4}}^2
-\mathbf{u}_{2}\mathbf{u}_{3}\eta_{\hat{4}}^2
+\mathbf{u}_{2}\mathbf{u}_{\hat{4}}\eta_{\hat{1}}^2
-\mathbf{u}_{2}\mathbf{u}_{3}\eta_{\hat{1}}^2
-\mathbf{u}_{\hat{1}}\mathbf{u}_{\hat{4}}\eta_{2}^2
+\mathbf{u}_{\hat{1}}\mathbf{u}_{3}\eta_{2}^2\Big) +\\
\nonumber \frac{\sqrt{s^{(6)}_{14}}}{2} & 
\Big( \mathbf{w}_{\hat{1}}\mathbf{u}_{2}\mathbf{w}_{3}\mathbf{u}_{\hat{4}}
+\mathbf{u}_{\hat{1}}\mathbf{w}_{2}\mathbf{w}_{3}\mathbf{u}_{\hat{4}}
+\mathbf{w}_{\hat{1}}\mathbf{u}_{2}\mathbf{u}_{3}\mathbf{w}_{\hat{4}}
+\mathbf{u}_{\hat{1}}\mathbf{w}_{2}\mathbf{u}_{3}\mathbf{w}_{\hat{4}}
\\
\nonumber & + \frac{1}{2} \left( 
\mathbf{w}_{\hat{1}}\mathbf{u}_{2}\eta_3^2+
\mathbf{u}_{\hat{1}}\mathbf{w}_{2}\eta_3^2-
\mathbf{w}_{\hat{1}}\mathbf{u}_{2}\eta_{\hat{4}}^2-
\mathbf{u}_{\hat{1}}\mathbf{w}_{2}\eta_{\hat{4}}^2+
\mathbf{w}_{3}\mathbf{u}_{\hat{4}}\eta_{\hat{1}}^2+
\mathbf{u}_{3}\mathbf{w}_{\hat{4}}\eta_{\hat{1}}^2-
\mathbf{w}_{3}\mathbf{u}_{\hat{4}}\eta_{2}^2-
\mathbf{u}_{3}\mathbf{w}_{\hat{4}}\eta_{2}^2 \right) \\
& \phantom{+} + \frac{1}{4}\left( \eta^2_3\eta^2_{\hat{1}}
+ \eta^2_2\eta^2_{\hat{4}}-\eta^2_3\eta^2_{2}- \eta^2_{\hat{4}}\eta^2_{\hat{1}} \right) \Big) 
\label{Half}
\end{align}
The claim is that the above is equivalent to: 
\eq
{\rm eq.~(\ref{Half})}=\frac{-1}{2\sqrt{-s^{(6)}_{14}}}\delta^4(Q),\;\;Q^A\equiv q_{\hat{1}}^A+q_{2}^A+q_{3}^A+q_{\hat{4}}^A
\label{Equivalent}
\eqe
To see this we can compare the explicit coefficients of the Grassmann polynomials of the two equations. The terms in eq.~(\ref{Equivalent}) evaluate to:
\begin{align}
\bullet~(\eta_{2a}\eta_{\hat{4}b}\eta^2_3): \qquad\qquad &
\frac{-1}{4\sqrt{-s^{(6)}_{14}}}\langle 2^a|\tilde{p}_3|\hat4^b\rangle=\frac{1}{4\sqrt{-s^{(6)}_{14}}}\langle 2^a|\tilde{3}_{\dot{a}}]\tilde{u}_{3}^{\dot{a}}u_{\hat4}^b=\frac{1}{4}u_2^au_{\hat{4}}^b \\
\bullet~(\eta^2_{2}\eta^2_3): \phantom{\eta_{2a}} \qquad\qquad & 
\frac{-1}{16\sqrt{-s^{(6)}_{14}}}p_2^{AB}\tilde{p}_{3AB}=\frac{-i\sqrt{-s^{(6)}_{14}}}{8}
\end{align}
These two terms agree with eq.~(\ref{Half}); the same follows for all others by similar arguments \cite{cheung0}. Given these results and using that in the BCFW supershift $q_{\hat{1}}+q_{\hat{4}}=q_{1}+q_{4}$, we find: 
\eq
{\rm eq.~(\ref{Integrate})}=\frac{-1}{4s^{(6)}_{14}}\delta^4(Q)\delta^4(\hat{Q})
\eqe

\end{document}